\documentclass{emulateapj}

\slugcomment{Submitted to the Astrophysical Journal}
\shorttitle{Clustering of $K$-selected Galaxies}
\shortauthors{Quadri et al.}

\begin{document}

\title{Clustering of $K$-selected Galaxies at $2<z<3.5$: Evidence for a Color-Density Relation}

\author{
Ryan Quadri\altaffilmark{1},
Pieter van Dokkum\altaffilmark{1},
Eric Gawiser\altaffilmark{1,2,3},
Marijn Franx\altaffilmark{4},
Danilo Marchesini\altaffilmark{1},
Paulina Lira\altaffilmark{2},
Gregory Rudnick\altaffilmark{5},
David Herrera\altaffilmark{1},
Jose Maza\altaffilmark{2},
Mariska Kriek\altaffilmark{4},
Ivo Labb\'{e}\altaffilmark{6,7},
Harold Francke\altaffilmark{1,2}
}

\altaffiltext{1}{Department of Astronomy, Yale University, New Haven, 
CT, PO Box 208101, New Haven, CT  06520}
\email{quadri@astro.yale.edu}
\altaffiltext{2}{Departamento de Astronom\'\i{}a, Universidad de Chile,
Casilla 36-D, Santiago, Chile}
\altaffiltext{3}{National Science Foundation Astronomy and Astrophysics
Postdoctoral Fellow}
\altaffiltext{4}{Leiden Observatory, PO Box 9513, NL-2300 RA,
Leiden, The Netherlands}
\altaffiltext{5}{National Optical Astronomical Observatory, 950 North
Cherry Avenue, Tucson, AZ 85719}
\altaffiltext{6}{Carnegie Observatories, 813 Santa Barbara Street, Pasadena,
CA 91101}
\altaffiltext{7}{Carnegie Fellow}

\begin{abstract}
  We study the clustering properties of $K$-selected galaxies at
  $2<z<3.5$ using deep multiwavelength imaging in three fields from
  the MUSYC survey.  These are the first measurements to probe the
  spatial correlation function of $K$-selected galaxies in this
  redshift range on large scales, allowing for robust conclusions
  about the dark matter halos that host these galaxies.  $K$-selected
  galaxies with $K<21$ have a correlation length $r_0 \sim 6 h^{-1}
  \textrm{Mpc}$, larger than typical values found for
  optically-selected galaxies.  The correlation length does not depend
  on $K$-band magnitude in our sample, but it does increase strongly
  with color; the $J-K>2.3$ distant red galaxies (DRGs) have $r_0 \sim
  11 h^{-1} \textrm{Mpc}$.  Furthermore, contrary to findings for
  optically-selected galaxies $K$-selected galaxies that are faint in
  the $R$-band cluster more strongly than brighter galaxies.  These
  results suggest that a color-density relation was in place at $z>2$;
  it will be interesting to see whether this relation is driven by
  galaxies with old stellar populations or by dusty star forming
  galaxies.  Irrespective of the cause, our results indicate that
  $K$-bright blue galaxies and $K$-bright red galaxies are
  fundamentally different, as they have different clustering
  properties.  Using a simple model of one galaxy per halo, we infer
  halo masses $\sim 5 \times 10^{12} M_\odot$ for $K<21$ galaxies and
  $\sim 2 \times 10^{13} M_\odot$ for DRGs.  A comparison of the
  observed space density of DRGs to the density of their host halos
  suggests large halo occupation numbers; however, this result is at
  odds with the lack of a strong small-scale excess in the angular
  correlation function.  Using the predicted evolution of halo mass to
  investigate relationships between galaxy populations at different
  redshifts, we find that the $z=0$ descendants of the galaxies
  considered here reside primarily in groups and clusters.
\end{abstract}

\section{Introduction}
\label{sec-intro}

Optical surveys of the high-redshift universe have been very
successful in finding relatively unobscured star-forming galaxies,
primarily via the $U$-dropout technique.  These $z \sim 3$ Lyman break
galaxies (LBGs) typically have stellar masses $\sim 10^{10} M_\odot$,
star formation rates of $10-100 M_\odot \textrm{yr}^{-1}$, and are
thought to dominate the star formation density at that epoch
\citep{steidel03,shapley01,reddy05}.  However, it is becoming
increasingly clear that substantial numbers of galaxies exist at these
redshifts that have little rest-frame UV luminosity and are thus
underrepresented in optical surveys.  Such galaxies may be detected in
the near-infrared (NIR), which samples the rest-frame optical.  

One criterion used to select galaxies in the NIR is $J-K>2.3$
\citep{franx03,vandokkum03}.  These distant red galaxies (DRGs)
typically have high star formation rates, $\gtrsim 100 M_\odot
\textrm{yr}^{-1}$, and dust obscuration $A_V > 1$, but must also have
significant populations of evolved stars in order to explain their
colors and spectra \citep{forster04,papovich06,kriek06a}.  Some DRGs
show little or no evidence of active star formation \citep{labbe05,
 kriek06a,kriek06b,reddy06}.  DRGs must in general be very massive to
account for their significant $K$-brightness; stellar population
synthesis models imply masses $M_* \sim ~10^{11} M_\odot$.  Indeed,
$~95\%$ of galaxies with $M_* > 10^{11} M_\odot$ at $2<z<3$ have
$K<21.3$ \citep{vandokkum06}.  Conversely, the median galaxy in this
mass range has $R \sim 25.9$, fainter than the limits typically
reached by optical surveys.

The relationship between $K$-selected samples and optically-selected
samples, not to mention present-day galaxies, remains unclear.  While
typical optically-selected galaxies have different properties than
$K$-selected galaxies, the $K$-bright subsample of optically-selected
galaxies have stellar masses, star formation rates, and metallicities
that are in approximate agreement with $K$-selected galaxies
\citep{shapley04,reddy05}.  Nevertheless, it is not clear whether the
differences between these galaxies are transient (e.g. dust geometry,
starbursts, mergers) or fundamental (e.g. age of the underlying old
stellar populations, mass, environment).  An understanding of the
nature of the differences between populations is essential to place
them in evolutionary scenarios.

One way to investigate differences between galaxy populations is to
measure their clustering properties.  As clustering measurements
provide information that is independent of photometric properties,
they can be used to distinguish between transient and fundamental
differences between galaxy populations.  In the halo model of galaxy
formation, the large-scale distribution of galaxies is determined by
the distribution of dark matter halos.  The correlation function of
galaxies can therefore be associated with the correlation function of
the halos in which they reside.  Halo clustering, in turn, is a strong
function of halo mass \citep{mo96}, providing a means to study the
relationship between galaxy properties and the mass of the dark matter
halos.  

Several studies have measured the dependence of clustering strength of
high-redshift galaxies on color.  \citet{daddi03} use the ultradeep
imaging of the $4.5 \textrm{arcmin}^2$ FIRES HDF-S field
\citep{labbe03} to study the clustering characteristics of
$K$-selected galaxies at $2<z<4$.  Their most striking finding is that
the correlation length increases strongly with $J-K$ color, with the
reddest galaxies in their sample having correlation lengths $r_0 =
10-15h^{-1} \textrm{Mpc}$, comparable to the most luminous red
galaxies in the local universe.  The mass of dark matter halos with
similar correlation lengths is $>10^{13} M_\odot$, yet the galaxy
number density is $\sim 100$ times larger than that expected for such
massive dark matter halos, implying that many galaxies must share the
same halo.  More recently, \citet{grazian06} measured a $r_0 =
13.4^{+3.0}_{-3.2} h^{-1} \textrm{Mpc}$ for $z>2$ DRGs using the
larger $135 \textrm{arcmin}^2$ GOODS-CDFS field, also indicating that
red galaxies are located in very massive halos.

The interpretation of these clustering measurements is complicated by
the fact that, in order to derive information about the dark matter
halos, the correlation function must be measured on large scales.  The
correlation function has a contribution from galaxies that share halos
(hereafter the ``1-halo'' term) and galaxies in separate halos (the
``2-halo'' term).  The shapes of these two contributions are shown
with impressive detail in the correlation functions of large samples
of $z \sim 4$ LBGs presented by \citet{lee06} and \citet{ouchi05}.  In
order to derive meaningful constraints on large-scale clustering
properties, and thus on the host dark matter halos, it is important
that both of these terms are taken into consideration.  If the
correlation function is parameterized as a simple power-law then it
should be measured on scales where the 2-halo term dominates the
clustering signal.  In practice, a firm lower limit to the radial
range in which the angular correlation function should be fitted is
the halo virial radius.  At $z=3$, the virial radius $r_{200}$ of a
$10^{13} M_\odot$ halo corresponds to $22 \arcsec$
\citep[e.g.][]{mo02}.  The majority of the clustering signal from
\citet{daddi03} and \citet{grazian06} is on scales $\theta \lesssim 30
\arcsec$, which may lead to gross overestimates of the large-scale
correlation length and the mass of the host dark matter halos.  In
particular, \citet{zheng04} shows that the measurements of
\citet{daddi03} are consistent with models in which the large-scale
correlation length is as low as $r_0 \sim 5 h^{-1}$ and the typical
halo mass is $\sim 10^{12} M_\odot$.  This suggests that DRGs and LBGs
may occupy similar halos, but that DRGs have higher occupation numbers.

The goal of this work is to study the clustering characteristics of a
$K$-selected population of galaxies at $2<z<3.5$.  The increased
field-of-view of our imaging allows for an improved determination of
the clustering strength of $K$-selected galaxies at angular
separations sufficiently large to investigate the large-scale
distribution of galaxies, and thus to provide more meaningful
estimates of the masses of the halos in which they reside.
Secondarily, we wish to analyze the clustering results using models of
halo clustering, and to use these models to shed light on evolutionary
scenarios for $z>2$ galaxies.  These types of analyses have previously
been performed for optically-selected samples
\citep[e.g.][]{moustakas02, ouchi04, adelberger05a}.  Throughout, we
use the cosmological parameters $\Omega_m=0.3$, $\Omega_\Lambda=0.7$,
$H_0=70 h_{70} \textrm{ km} \textrm{ s}^{-1} \textrm{ Mpc}^{-1}$, and
$\sigma_8=0.9$.  Results are given using $h_{70}=1$, except for
correlation lengths which are scaled to units of $h=0.7$ in order to
facilitate comparison to previous studies.  Optical magnitudes are
given in AB and NIR magnitudes are given on the Vega system.

\section{Data}
\label{sec:data}

The Multiwavelength Survey by Yale-Chile (MUSYC) consists of optical
and NIR imaging of four independent $30\arcmin \times 30\arcmin$
fields plus spectroscopic follow-up \citep[][R. Quadri et al. 2006, in
preparation]{gawiser06}.\footnote{www.astro.yale.edu/MUSYC} Deeper
$JHK$ imaging was obtained over $10\arcmin \times 10\arcmin$
sub-fields with the ISPI camera at the CTIO Blanco 4m telescope.  The
present analysis is restricted to three of these deep fields (the
adjacent HDFS1 and HDFS2, and SDSS 1030).  The deep $JHK$ data will be
described in detail elsewhere (R. Quadri et al. 2006, in preparation).
The total $5\sigma$ point source limiting depths are $J \sim 23.0$, $H
\sim 21.8$, and $K \sim 21.3$.  The optical $UBVRIz$ data are
described in \citet{gawiser06}.

\begin{figure}
  \epsscale{1.1}
  \plotone{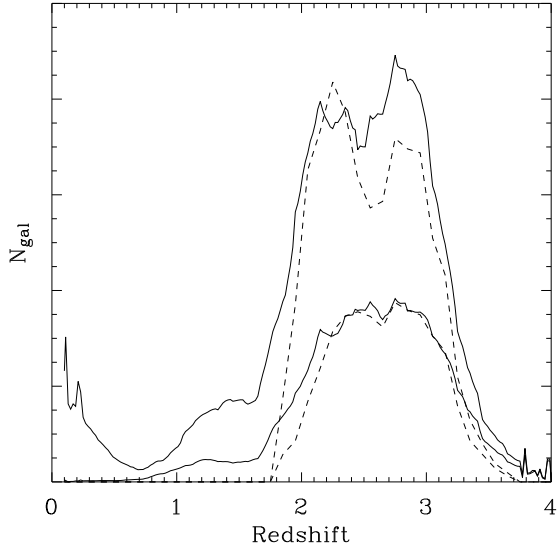}
  \caption{The inferred redshift distribution of galaxies selected with
    $2 < z_{\rm{phot}} < 3.5$.  The upper two curves are for the full sample,
    and the lower two curves are for galaxies that meet the $J-K>2.3$
    criterion for DRGs.  Dashed curves indicate the distribution of
    $z_{\rm{phot}}$ values, smoothed with a $\Delta(z)=0.4$ boxcar average.
    Solid curves indicate the distributions derived by summing the
    redshift probability distributions for each galaxy.  The
    normalization is arbitrary.}
  \label{fig:redshift_dist}
\end{figure}

\begin{figure*}
  \begin{center}
    \epsscale{1.1}
    \plotone{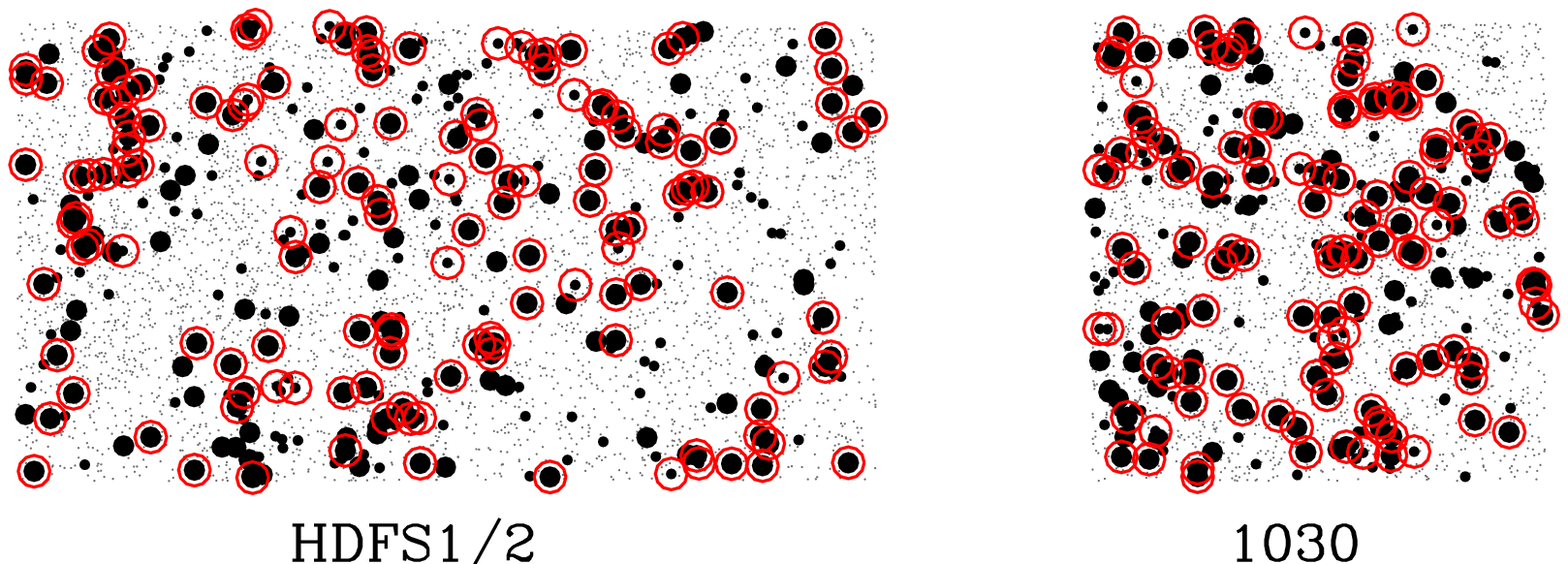}
    \caption{The position of $K<21$ galaxies at $2<z_{\rm{phot}}<3.5$
      in the deep MUSYC fields.  The field sizes are $\sim 20\arcmin
      \times 10\arcmin$ and $\sim 10\arcmin \times 10\arcmin$ for
      HDFS1/2 and 1030, respectively.  The large black circles
      represent galaxies with stellar mass $M_*>10^{11}M_\odot$ and
      the small black circles represent less massive galaxies.
      Galaxies that meet the $J-K>2.3$ criterion for distant red
      galaxies are marked with an open red circle.}
    \label{fig:xy}
  \end{center}
\end{figure*}

In this study we use spectroscopic redshifts where possible, but must
rely primarily on photometric redshifts.  Photometric redshifts were
determined using the methods of \citet{rudnick01,rudnick03}.  Briefly,
non-negative linear combinations of galaxy templates are fit to the
observed spectral energy distributions.  The templates include the
four empirical templates of \citet{coleman80}, as well as the two
empirical starburst templates of \citet{kinney96}, all of which have
been extended into the UV and NIR using models.  As the empirical
templates are derived from low-redshift samples, we find that they do
not adequately describe all $z \gtrsim 2$ galaxies.  For this reason
we added $10\textrm{Myr}$ and $1\textrm{Gyr}$ old single stellar
population templates generated with the \citet{bruzual03} models.
The redshift probability distribution for each galaxy is calculated
using Monte Carlo simulations in which the observed fluxes are varied
within the photometric uncertainties.

A comparison of the photometric redshifts to spectroscopic redshifts
drawn from the literature and from our own observations yields a mean
$\Delta z / {(1+z)} = 0.12$ for $z > 1.5$, corresponding to $\Delta z
\sim 0.4$ at $z \simeq 2.5$.  The dashed curves in
Fig.~\ref{fig:redshift_dist} show the redshift distribution for all
MUSYC galaxies with $2<z<3.5$, and for DRGs in the same redshift
range.  The solid curves show the distributions that are inferred by
summing the redshift probability distributions.  All distributions
have been smoothed with a $\Delta z=0.4$ boxcar to limit spikes that,
given our uncertainties, may not be real.

We restrict the sample to galaxies with $2<z_{\rm{phot}}<3.5$ and
$K<21$, except where noted.  Fig.~\ref{fig:xy} shows the positions of
the DRGs in the three MUSYC fields.  Also shown are galaxies with
stellar mass $M_*>10^{11} M_\odot$, where $M_*$ is determined with
stellar population synthesis models (\S \ref{sec:stellarmass}).

\section{The two-point correlation function}
\label{sec-corrfxn}

\subsection{Method}
\label{sec:method}

The two-point correlation function can be measured by counting the
number of unique galaxy pairs as a function of separation, and
comparing the resultant distribution to that of a catalog of random
points with the same number density and subject to the same observing
geometry.  Several estimators for the angular two-point correlation
function are available, but the estimator introduced by
\citet{landy93} is emerging as the de facto standard for high-redshift
studies.  It has been shown to minimize the variance and biases
associated with other estimators
\citep{landy93,hamilton93,kerscher00}.  The observed amplitude of the
two point correlation function is thus
\begin{equation}
w_{obs}(\theta) = \frac{DD(\theta)-2DR(\theta)+RR(\theta)}{RR(\theta)},
\label{eq:landyszalay}
\end{equation}
where $DD(\theta)$ is the number of data-data pairs with angular
separation in the interval
$(\theta-\Delta\theta/2,\theta+\Delta\theta/2$).  $DR(\theta)$ is the
number of data-random pairs, and $RR(\theta)$ is the number of
random-random pairs, in the same angular interval.  We use
$\Delta\theta = 20\arcsec$.  In order to better sample the observing
geometry, and to decrease the uncertainty in $DR(\theta)$ and
$RR(\theta)$, we use $\sim 100$ times more random points than data
points.  This requires normalizing the $DR$ and $RR$ terms such that
$\sum_\theta DR(\theta) = \sum_\theta RR(\theta) = \sum_\theta
DD(\theta)$.

The angular correlation function can be approximated as a power law
\begin{equation}
w(\theta) = A_w\theta^{-\beta}.
\end{equation}
However, as the (suitably normalized) number of random pairs is equal to
the number of data pairs, and since the two-point correlation
function is the \emph{excess} probability of finding a data pair
versus a random pair, it is clear that $w_{obs}(\theta)$ cannot be
positive for all $\theta$.  In particular,
\begin{equation}
\int \int w_{obs}(\theta_{12}) d\Omega_1 d\Omega_2 \approx 0.
\end{equation}
This integral constraint requires that $w_{obs}(\theta)$ fall below
the intrinsic $w(\theta)$ \citep{groth77}.  The size of this bias
increases with the clustering strength and decreases with field size;
in practice, it is a significant effect and a correction must be made.
The integral constraint correction is approximately constant and equal
to the fractional variance of galaxy counts in a field,
\begin{equation}
IC \approx \sigma^2 = \frac{1}{<N_{gal}>} + \sigma_w^2,
\label{eq:sigmasq}
\end{equation}
where the first term on the right is the Poisson variance and the
second accounts for the additional variance caused by clustering
\begin{equation}
\sigma_w^2 = \frac{1}{\Omega^2} \int \int w(\theta_{12}) d\Omega_1 d\Omega_2
\end{equation}
\citep[][\S 45]{peebles80}.  Although the clustering term dominates
the integral constraint, the Poisson term is non-negligible for the
small sample sizes considered here.  Following \citet{infante94} and
\citet{roche99} the clustering term $\sigma_w^2$ can be estimated
numerically using
\begin{equation}
\sigma_w^2 = \frac{ \sum_i A_w\theta_i^{-\beta} RR(\theta_i) } { \sum_i RR(\theta_i) }.
\label{eq:sigmasqnum}
\end{equation}
The quantity $\sigma_w^2/A_w$ is estimated directly from the random
catalog for an assumed value $\beta$.  The amplitude $A_w$ of the
angular correlation function is related to the observations through
the fitting function
\begin{equation}
w_{obs}(\theta) = A_w\theta^{-\beta} - IC.
\label{eq:w}
\end{equation}
We estimate $A_w$ iteratively using eqs. \ref{eq:sigmasqnum},
\ref{eq:sigmasq}, and \ref{eq:w}.  The final result is robust against
differences in the initial estimate of $A_w$ and convergence only
takes a few iterations.

In the weak clustering regime the uncertainty in the Landy \& Szalay
estimator can be estimated by assuming that $DD(\theta)$ has Poisson
variance \citep{landy93}; in this case
\begin{equation}
\delta w_{obs}(\theta) \approx \frac{1+w(\theta)}{\sqrt{DD(\theta)}}.
\end{equation}

If the angular correlation function is a power law, the spatial
correlation function will also be a power law
\begin{equation}
\xi(r) = \left(\frac{r}{r_0}\right)^{-\gamma},
\label{eq:xi}
\end{equation}
where $r_0$ is the spatial correlation length and ${\gamma=\beta+1}$.
The angular correlation function can be used to obtain the spatial
correlation function by inverting the Limber projection
\begin{equation}
A_w = \frac { H_\gamma r_0^\gamma \int F(z) r_c^{1-\gamma}(z) N^2(z) E(z) dz } {(c/H_o)(\int N(z) dz)^2},
\label{eq:limber}
\end{equation}
where $r_c(z)$ is the comoving radial distance, N(z) is the redshift
distribution,
\begin{equation}
H_\gamma = \Gamma(1/2) \frac {\Gamma[(\gamma-1)/2]} {\Gamma(\gamma/2)},
\end{equation}
and
\begin{equation}
E(z) = \sqrt{\Omega_m(1+z)^3 + \Omega_\Lambda},
\end{equation}
\citep[e.g.][]{magliocchetti99}.

The function $F(z)$ describes the evolution of clustering with
redshift, $\xi(r,z) = \xi(r,0)F(z)$.  The evolution has often been
modelled as $F(z)=(1+z)^{-(3-\gamma+\epsilon)}$, where the parameter
$\epsilon$ is typically specified using $\epsilon=\gamma-3$ for
constant clustering in comoving units, $\epsilon=0$ for `stable
clustering', or $\epsilon=\gamma-1$ for `linear growth'
\citep[e.g.][]{moscardini98, overzier03}.  We assume constant
clustering in comoving units over $2<z<3.5$; this sets $F(z)=1$.
Different values of $\epsilon$, where the correlation length is then
determined at the median redshift of the observed sample, yield
similar results.


\subsection{Measurement Strategies}

In what follows we restrict the analysis to galaxies with photometric
redshift $2 < z_{\rm{phot}} < 3.5$.  Reducing the redshift range
produces comparable correlation lengths, but with larger
uncertainties.  Additionally, as we largely rely on photometric
redshifts, we cannot be confident in our ability to divide the sample
too finely in redshift space.

It is common practice in the literature to assume $\gamma=1.8$ if the
data are not sufficient to make independent measurements of both the
slope and the amplitude of the correlation function.  Recent studies
have found that $\gamma \sim 1.6$ may be more appropriate for LBGs
\citep{adelberger05a, lee06}.  Direct comparisons of the correlation
length from different studies can be problematic unless the same
$\beta$ was used; for this reason, the results summarized below use
$\gamma=1.8$ but Tables~\ref{tbl:results_allrad} and \ref{tbl:results}
also gives the $r_0$ values corresponding to $\gamma=1.6$.

In placing the random objects on the image, we mask out regions where
galaxies could not be detected, e.g. in the vicinity of bright stars.
This procedure makes a negligible change in the resultant correlation
functions.

We measure $w_{obs}(\theta)$ (eq.~\ref{eq:landyszalay}) in
linearly-spaced $20 \arcsec$ bins for the purpose of computing the
$\chi^2$ fits.  Spacing the bins at equal logarithmic intervals gives
similar results for most samples considered here.  We present the
results of power law fits over the range $0\arcsec < \theta <
200\arcsec$ in Table~\ref{tbl:results_allrad}, and $40\arcsec < \theta
< 200\arcsec$ in Table~\ref{tbl:results}.  Most of the discussion and
analysis in this paper is based on the latter fits.  The $200 \arcsec$
upper limit minimizes the effects of low level biases (such as errors
in the flat-fielding and the integral constraint) and edge effects.
The $40\arcsec$ lower limit is set so that the fit is not strongly
affected by 1-halo term of the correlation function (\S
\ref{sec:halos}).  At $z=2$, the minimum redshift of our sample,
$40\arcsec$ subtends $0.7h^{-1} \textrm{Mpc}$ in comoving units, which
corresponds to the virial radius $r_{200}$ of an $\sim 3 \times
10^{13} M_\odot$ halo.  This is roughly the mass of halos that host
the most clustered galaxies in our sample (\S \ref{sec:halos}), and is
larger than the scales ($<10\arcsec$; $<0.25h^{-1} \textrm{Mpc}$) at
which $ z\sim 4$ LBGs show significant contributions from the 1-halo
term \citep{ouchi05, lee06}.  For reference, \citet{adelberger05a}
uses a lower limit of $60\arcsec$, but our reduced signal-to-noise
does not allow for such a conservative limit.  In contrast,
\citet{grazian06} fit $w(\theta)$ over $1\arcsec < \theta <
100\arcsec$.  Neither \citet{grazian06} nor \citet{daddi03}
significantly constrain $w(\theta)$ beyond $40\arcsec$ (see their
Figs. 9 and 8, respectively).

We note that, for most samples considered here, the
$0\arcsec<\theta<200\arcsec$ fits result in correlation lengths that
are larger than the $40\arcsec<\theta<200\arcsec$ correlation lengths
by $\sim 1-1.5 \sigma$.  For some subsamples the
$0\arcsec<\theta<200\arcsec$ correlation lengths are actually smaller,
although the difference is always $\leq 1 \sigma$.

While performing the fit at large scales reduces the effect of the
1-halo term on the correlation length, there is a second-order effect
of the halo occupation distribution that we do not take into account.
A fully consistent treatment would require counting only one galaxy
per halo, to avoid counting the same halo more than once.  As we have
no robust method to detect galaxies that share halos, halos that host
multiple galaxies will be counted multiple times when measuring
$w(\theta)$.  Since only the most massive halos are likely to host
multiple galaxies, these halos effectively receive more weight.  Our
data are not sufficient to address these second-order effects, and we
note that simply rejecting all galaxies that have close neighbors
would introduce other biases in our measurements.

\subsection{Sources of uncertainty}
\label{sec:uncertainties}

\subsubsection{Redshift distribution}

The shape of the redshift distribution $N(z)$ in eq.~\ref{eq:limber}
will affect the deprojection of the angular two point correlation
function, contributing to the uncertainty in $r_0$.  One strategy for
dealing with the redshift uncertainties is to smooth the
$z_{\rm{phot}}$ distribution by the typical uncertainty, $\Delta z
\sim 0.4$ (\S \ref{sec:data}).  However, it is likely that the
redshift uncertainties differ for different galaxies in our sample.
This may affect the observed relationships between galaxy properties
and clustering strength; for instance, if faint galaxies have a larger
photometric redshift uncertainty than bright galaxies, their intrinsic
redshift distribution may be wider.  In this case, smoothing with the
same $\Delta z$ will not sufficiently broaden the redshift
distribution of faint galaxies, resulting in an underestimate of the
correlation length of faint galaxies for some observed value of $A_w$.
This will introduce an artificial trend of increasing clustering with
increasing brightness.  We note that many other studies of galaxy
clustering--which use the same redshift selection function for all
galaxy samples--may be subject to this effect.

A more appropriate redshift distribution for use in
eq.~\ref{eq:limber} may be had by summing the redshift probability
distribution $P(z)$ (see \S \ref{sec:data}) for each galaxy in the
sample.  To the extent that our Monte Carlo simulations provide an
accurate estimate of $P(z)$, this strategy circumvents the problem of
choosing a smoothing width.  Additionally, we compute $\sum P(z)$
separately for each sub-sample under consideration, thereby reducing
the problem of differential redshift uncertainties.

The smoothed $z_{\rm{phot}}$ distributions are narrower than the (more
realistic) $\sum P(z)$ distributions that are used throughout this
paper.  We note that using these narrower distributions would
reduce our estimates of $r_0$ by $\sim 15-25\%$, where the more
clustered populations display the smaller differences.  In the case of
DRGs, about half of the $\sim 15\%$ difference between these two
estimates of $r_0$ comes from the $z < 1.8$ tail of the $\sum P(z)$
distribution, and the other half comes from the broader overall
distribution at $z > 1.8$ (Fig.~\ref{fig:redshift_dist}).


If redshift interlopers are assumed to be randomly distributed they
will dilute observed angular correlation by a factor $(1-f_c)^2$,
where $f_c$ is the contamination fraction.  One method to account for
interlopers is to estimate $f_c$ using the redshift probability
distributions, and to calculate a contaminant-corrected $r_0$ by
integrating over the range $1.8<z<3.5$ in eq.~\ref{eq:limber}.  As the
assumption of a random distribution is probably unrealistic, we choose
instead to account for interlopers by integrating over the entire
redshift range.  The correlation lengths are similar regardless of the
method used.

\subsubsection{Field to Field Variance}
\label{sec:simulations}

The Landy \& Szalay estimator has been shown to have approximately
Poisson variance in the limit of zero clustering \citep{landy93}, but
a clustered population is expected to show covariance between the
radial bins \citep{bernstein94}.  The deep NIR MUSYC survey consists
of only two independent fields (HDFS1/2 and 1030), so field to field
variations will present an additional source of error.  Moreover,
uncertainty in the integral constraint is not correctly accounted for
by Poisson statistics.  Here we estimate the confidence intervals of
our results with simulated data sets.  Our approach is similar in
spirit to that described by \citet{daddi03}.

We construct a clustered population using outputs from the public
GalICS simulations \citep{hatton03}.  GalICS uses cosmological N-body
simulations to trace the growth and merging of dark matter halos, and
a semi-analytic approach to follow the formation and evolution of
galaxies within the halos.  The simulations use $8.272 \times 10^{9}
M_\odot$ particles in a $100h^{-1} \textrm{Mpc}$ simulation
box, $\Omega_m = 0.333$, $\Omega_\Lambda = 0.667$, $h = 0.667$, and
$\sigma_8 = 0.88$.  The GalICS outputs are available in convenient
`observing cone' catalogs \citep{blaizot05} that mimic what an
observer would see in a simulated universe.  The limited size of the
simulation box requires replication of galaxies within the observing
cones.  Although the observing cone geometry has been tuned to reduce
replication effects, precise measurements of galaxy clustering and the
cosmic variance are hampered by replication effects.

We construct $27$ mock data sets from the eight $1 \textrm{deg}^2$
GalICS observing cones.  Each of these mock data sets has the same
geometry and field sizes as the deep NIR MUSYC survey, and we measure
the clustering of the simulated galaxies using the same methods.  The
results are used to estimate the $68\%$ confidence range of the
amplitude $A_w$ of the angular correlation function.  More detailed
characterizations of the uncertainties would require larger
simulations.  The confidence range is a function of both intrinsic
clustering, which is adjusted by selecting galaxies with different
halo mass, and surface density, which is adjusted by randomly removing
galaxies.  For most of the samples here, $\sigma_{A_w} / A_w \approx
55\% - 70\%$--significantly larger than the Poisson values alone.
This does not imply that our results are only significant at the
$\lesssim 2\sigma$ level, as populations with small $A_w$ will have
small $\sigma_{A_w}$, and so populations with little or no clustering
rarely show strong clustering \citep[see also Fig.~1 of][]{daddi03}.
We will return to this issue in \S \ref{sec:clustcolor} for the
specific case of DRG vs. LBG clustering.

The errors due to field to field variations are reduced when comparing
populations of galaxies drawn from the same fields, so we quote
Poisson uncertainties except where noted.  The estimated total
uncertainty in the correlation length due to both Poisson errors and
field to field variance is given in the last column of
Table~\ref{tbl:results}.  We note that most studies of galaxy
clustering at high redshift do not fully account for the effects of
field to field variance.

\subsection{Results}
\label{sec:clustresults}

\subsubsection{Angular and spatial clustering of galaxies with $2<z_{\rm{phot}}<3.5$}

\begin{figure}
  \epsscale{1.1}
  \plotone{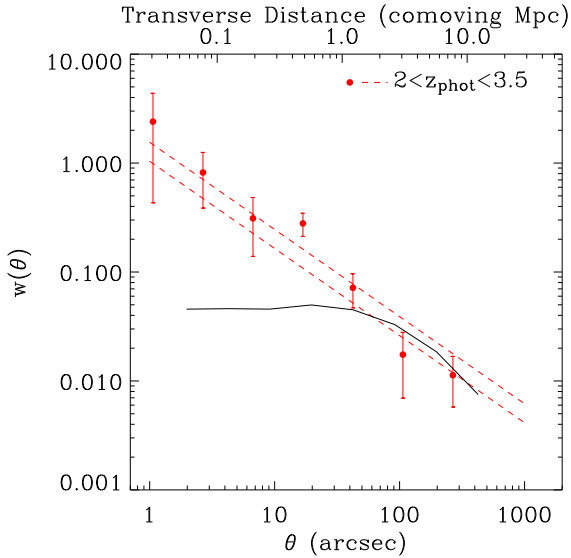}
  \caption{The angular correlation function, with the integral
    constraint correction applied to the data points, for galaxies
    with $2 < z_{\rm{phot}} < 3.5$ and $K<21$.  The upper x-axis shows
    the transverse comoving distance at the median redshift $z=2.6$.
    The dashed lines show the best fit power laws over $\theta <
    200\arcsec$ (upper) and $40\arcsec < \theta < 200\arcsec$ (lower).
    The lower fit is preferred in order to reduce the effects of halo
    substructure in $w(\theta)$.  The solid curve illustrates the
    shape of the angular correlation function that is expected for
    dark matter halos; the larger values of the galaxy correlation
    function at small separations suggest that some halos host
    multiple galaxies.}
  \label{fig:wK21}
\end{figure}

Fig.~\ref{fig:wK21} shows the angular correlation function for
$K$-selected galaxies with $2<z_{\rm{phot}}<3.5$ and $K<21$.  The
correlation function is roughly consistent with a power law down to
$\sim 1 \arcsec$, the approximate resolution limit of our survey.
There is a slight indication of an excess on smaller scales: the
amplitude of the best-fitting power law $A_w = 1.9 \pm 0.3$ when the
fit is restricted to $\theta < 200\arcsec$ and $A_w = 1.3 \pm 0.4$
over $40\arcsec < \theta < 200\arcsec$.  \citet{lee06} and
\citet{ouchi05} also note that the best-fitting power law changes for
smaller angular intervals for their sample of $z \sim 4$ LBGs.
Although our sample is not large enough to trace the detailed shape of
$w(\theta)$, it is possible that this is evidence of the small-scale
excess that is predicted by halo occupation distribution models
\citep[e.g.][]{wechsler01,zheng04}.  However, as emphasized by
\citet{adelberger05a}, the observation that approximate power law
behavior extends to such small scales may itself be interpreted as
evidence that galaxies share halos.  The solid line in
Fig.~\ref{fig:wK21} shows the expected shape of the angular
correlation function of dark matter halos $w_h(\theta)$, derived using
N-body simulations (\S \ref{sec:simulations}).  Halo exclusion effects
force $w_h(\theta)$ to flatten on smaller scales.  In the case of one
galaxy per halo, $w(\theta)$ should follow $w_h(\theta)$.  If galaxies
have a higher probability than halos of having close neighbors, then
it is likely that some fraction of these galaxy neighbors reside in
the same halo.  We return to this point in \S \ref{sec:halos}.

We invert the angular correlation function $w(\theta)$ to derive the
spatial correlation length $r_0$ using eq.~\ref{eq:limber}.
Restricting the fit to the angular range $40\arcsec < \theta <
200\arcsec$, we find $r_0 = 6.0^{+0.9}_{-1.1} h^{-1} \textrm{Mpc}$
(comoving).  For comparison, the correlation lengths of $R<25.5$
optically-selected $z \sim 2-3$ BX galaxies and LBGs is $\sim 4 h^{-1}
\textrm{Mpc}$ \citep{adelberger05a, lee06}.  The latter use a power
law slope of the correlation function $\gamma \simeq 1.6$, whereas we
use $\gamma = 1.8$.  Assuming $\gamma = 1.6$ increases the correlation
length of MUSYC galaxies to $r_0 = 6.6^{+1.0}_{-1.1} h^{-1}
\textrm{Mpc}$.  The larger correlation lengths for $K$-selected sample
might have been expected, as $K$-bright galaxies have been shown to
cluster more strongly than optically-selected $K$-faint galaxies at $z
\sim 2$ \citep{daddi04,adelberger05b}.  This dependence on selection
filter may reflect underlying trends with $K$-magnitude, color, mass,
or other parameters; these issues are discussed in the following
subsections (see particularly \S\ref{sec:magnitude}).

\subsubsection{Clustering as a function of color}
\label{sec:clustcolor}

\begin{figure}
  \epsscale{1.1}
  \plotone{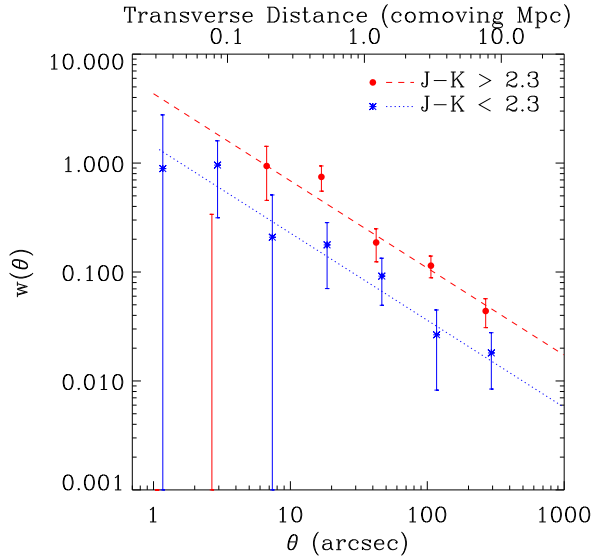}
  \caption{The angular correlation function for $2 < z_{\rm{phot}} <
    3.5$ galaxies that meet the $J-K > 2.3$ threshold for distant red
    galaxies (DRGs), and for non-DRGs.  A small horizontal offset has
    been applied to the blue points for display purposes only.  The
    power law fits to $w(\theta)$ are performed at $40\arcsec < \theta
    < 200\arcsec$ to reduce the effects of halo substructure in
    $w(\theta)$.}
  \label{fig:wJK}
\end{figure}

Fig.~\ref{fig:wJK} compares the angular correlation function of DRGs
with $J-K>2.3$ to that of non-DRGs with $J-K<2.3$, in the redshift
range $2 < z_{\rm{phot}} < 3.5$ and with $K<21$.  The DRGs cluster
more strongly than the bluer galaxies at large scales.  The angular
correlation function of DRGs is roughly consistent with a power law
down to $\sim 4 \arcsec$, which corresponds to the virial radius of a
$\sim 10^{11} M_\odot$ halo.  For the blue galaxies, $w(\theta)$
remains consistent with a power law to $\sim 1 \arcsec$.  The two
$w(\theta)$ data points at $\theta < 4 \arcsec$ correspond to $1$ and
$8$ observed blue galaxy pairs respectively, whereas the DRGs have $0$
and $2$ pairs at these small separations.  Extrapolating the power law
fit to these scales, we would expect only $1$ and $4$ galaxy pairs for
the DRG sample.  While galaxy pairs at separations approaching
$40\arcsec$ may share the same halo, it is not clear whether or not
they will eventually merge.  However galaxies with these significantly
smaller separations ($4\arcsec$ corresponds to a projected distance of
$36$ proper kpc at $z=2.6$) should be interacting strongly and could
be mergers in progress.  The lack of close DRG pairs may therefore
indicate that DRGs are undergoing few mergers.  Whether this is
because DRGs are the products of recent mergers, or because
merger-induced star formation makes the red galaxies bluer, or whether
there is some other explanation, is unknown.  Evaluating the
significance of the lack of close DRG pairs is severely complicated by
galaxy deblending issues in our $\sim 0.9\arcsec$ FWHM images, and we
will not discuss this issue further.
We find $r_0 = 11.1^{+1.3}_{-1.4}h^{-1} \textrm{Mpc}$ for DRGs.  Note
that, if the full field to field variance is taken into account, we
estimate $r_0 = 11.1^{+2.8}_{-4.2}h^{-1} \textrm{Mpc}$.  If the fit is
performed over $\theta<200\arcsec$, then $r_0 = 12.0^{+0.9}_{-1.0}
h^{-1} \textrm{Mpc}$ (Poisson errors only).  These values are
consistent with the $r_0 = 13.4^{+3.0}_{-3.2} h^{-1} \textrm{Mpc}$
given by \citet{grazian06} and $r_0 = 14.5^{+3.1}_{-3.7} h^{-1}
\textrm{Mpc}$ given by \citet{daddi03}, although the latter authors do
not apply a photometric redshift cut.  

\begin{figure}
  \epsscale{1.1}
  \plotone{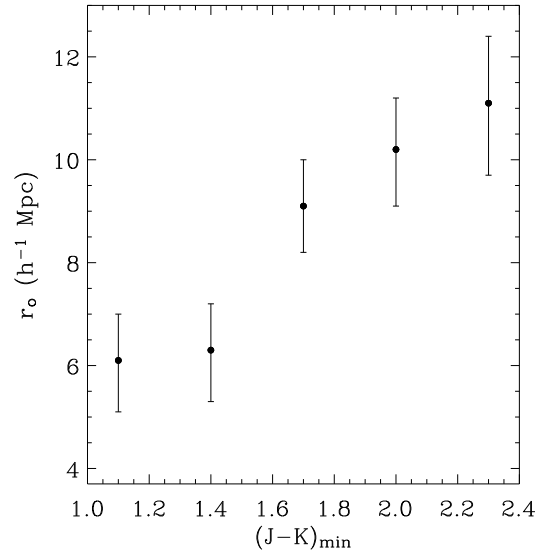}
  \caption{The comoving correlation length for $2 < z_{\rm{phot}} < 3.5$
    galaxies redder than the $J-K$ color threshold.}
  \label{fig:rJK}
\end{figure}

\begin{figure}
  \epsscale{1.1}
  \plotone{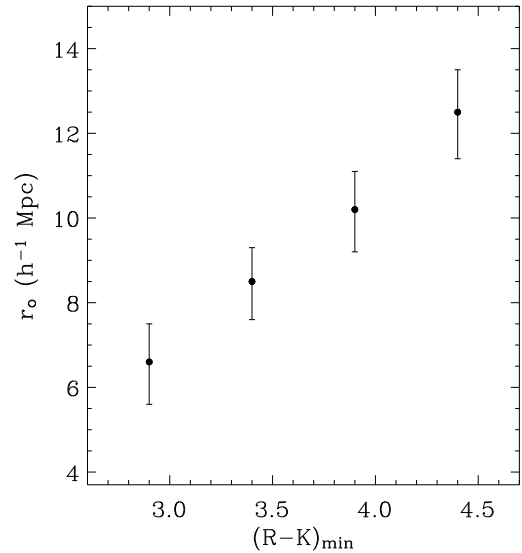}
  \caption{The comoving correlation length for $2 < z_{\rm{phot}} < 3.5$
    galaxies redder than the $R_{AB}-K_{Vega}$ color threshold.}
  \label{fig:rRK}
\end{figure}

Fig.~\ref{fig:rJK} shows the comoving correlation length as a function
of minimum $J-K$ color threshold.  We confirm the previous result of
\citet{daddi03} that the redder galaxies cluster more strongly, even
though their result was derived using a single $\sim 4.5 \textrm{
  arcmin}^2$ field, and they measure $w(\theta)$ at $\theta \lesssim
70\arcsec$ (see their Fig. 8).  It should also be noted that their
sample reaches $2-3$ magnitudes deeper than ours, and it is not
obvious that the same trends should hold over such a wide luminosity
range.  There is a slight trend of increasing median redshift with
increasing $J-K$ color, but the difference is $\approx 0.1$ over the
range of colors studied here, so it is unlikely that the relationship
between color and $r_0$ is solely due to redshift evolution.  We note
that alternate galaxy colors, such as $R-K$, are also strongly
correlated with clustering (Fig.~\ref{fig:rRK}).  LBGs and BX galaxies
have a correlation length $r_0 \approx 4 h^{-1} \textrm{Mpc}$
\citep{adelberger05a, lee06}, lower than the value for the bluest
threshold shown in Fig.~\ref{fig:rJK}, although the brightest $R<24$
LBGs reach $r_0 = 7.8 \pm 0.5 h^{-1} \textrm{Mpc}$ \citep{lee06}.  The
median $J-K$ color of $z \sim 3$ LBGs is $\sim 1.6$ \citep{shapley01},
and very few LBGs/BX galaxies reach the reddest thresholds
\citep{reddy05,vandokkum06}.

We have established the significance of the increased clustering with
color in several ways.  Splitting our sample at the median color, $J-K
\sim 2.17$, we find $r_0 = 11.0^{+1.1}_{-1.2} h^{-1} \textrm{Mpc}$ and
$r_0 = 6.1^{+1.8}_{-2.5} h^{-1} \textrm{Mpc}$ for the redder and bluer
sample, respectively.  We then randomly split the sample of
$K$-selected galaxies in two repeatedly, measuring the correlation
length each time.  The correlation length reaches as high as $r_0 = 11
h^{-1} \textrm{Mpc}$ only $\sim 4\%$ of the time, indicating that we
have established the stronger clustering for redder samples at the
$\sim 96\%$ level.  We have also used the simulations described in \S
\ref{sec:simulations} to see how often a population with the same
correlation length as LBGs, but number density and redshift
distribution similar to what we infer for DRGs, can have a measured
correlation length as high as that observed for DRGs as a result of
field to field variations; we found that this only happens $\sim 5\%$
of the time.  Futhermore, we have verified that the increase in
clustering with color is not driven by any one of our three ISPI
fields by repeating the clustering measurements three times, each time
removing one of the fields; although the exact values of the
correlation length vary, the relationship between clustering and color
is always present.  Finally, we recall that the `total' uncertainties,
which include the estimated contribution from field to field
variations and are presented in Table~\ref{tbl:results}, overestimate
the uncertainties when comparing correlation lengths of galaxies that
are drawn from the same fields.

The increasing clustering with color indicates that a color-density
relationship was in place at $z \gtrsim 2$.  In the local universe,
this relationship is understood as an effect of higher metallicity and
higher stellar ages in the densest regions; both effects may play a
role at high redshift \citep{forster04,vandokkum04,shapley04}.  The
high dust obscuration associated with vigorous starbursts also
contributes to the red colors of many $K$-selected galaxies
\citep[e.g.][]{forster04,labbe05,webb06}.  It is entirely possible
that the dusty and the ``red and dead'' galaxies \citep{kriek06b} have
different clustering properties, but the strong relationship between
$J-K$ and $r_0$ suggests that neither of these populations is weakly
clustered.  Disentangling the relationship between clustering and star
formation for red galaxies would likely require large fields with
\emph{Spitzer Space Telescope} observations \citep{labbe05}.

\subsubsection{Clustering as a function of apparent magnitude}
\label{sec:magnitude}

\begin{figure}
  \epsscale{1.1}
  \plotone{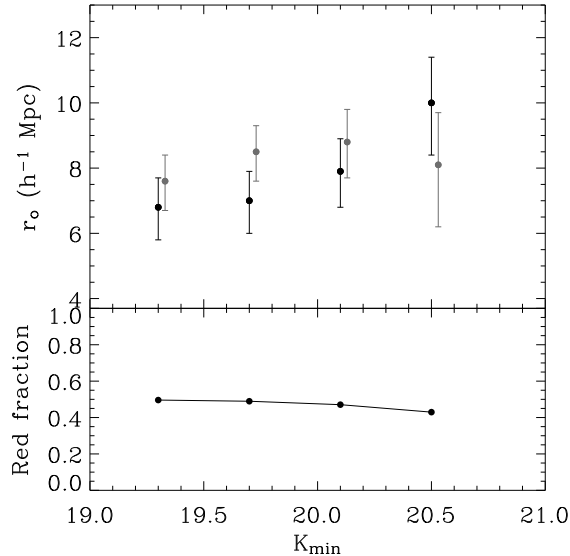}
  \caption{The upper plot shows the comoving correlation length for
    galaxies fainter than a $K$-magnitude threshold.  The grey points
    show the results from the HDFS1 and 1030 fields only; these points
    have been offset in the horizontal direction for clarity.  The
    lower plot shows that the fraction of red galaxies (defined as
    galaxies redder than the median, $J-K \geq 2.17$) does not change
    strongly over this magnitude range.}
  \label{fig:rK}
\end{figure}

Fig.~\ref{fig:rK} shows the relationship between correlation length
and minimum $K$-magnitude.  There is a small trend with the fainter
galaxies clustering more strongly, but the significance of this effect
is low; removing either the HDFS1 or HDFS2 fields from this analysis
eliminates this relationship, while removing 1030 actually increases
it.  We conclude that the data do not suggest a strong relationship
between $K$ and $r_0$.  This contrasts with results from $z \sim 2.3$
BX objects, which show clustering that increases strongly with $K$
\citep{adelberger05b}; we comment further on this below.  The bottom
panel of Fig.~\ref{fig:rK} shows that the fraction of red
galaxies---where here we characterize a galaxy as red if it has $J-K$
larger than the median---does not vary significantly with $K$.
Combined with the result that $r_0$ correlates with $J-K$, it appears
that color, and not $K$-magnitude, is the primary determinant of
clustering strength.

\begin{figure}
  \epsscale{1.1}
  \plotone{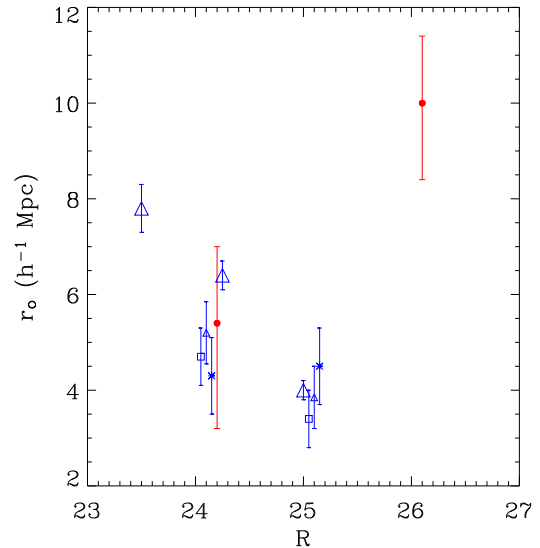}
  \caption{The comoving correlation length for $R$-selected (blue
    symbols) and $K$-selected galaxies (red circles), shown at
    representative $R$-magnitude values.  The large triangles are for $z
    \sim 3$ LBGs from \citet{lee06}.  The small triangles are LBGs at
    $z \sim 3$, the squares are BX objects at $z \sim 2.3$, and the
    asterisks are BM objects at $z \sim 1.7$, all taken from
    \citet{adelberger05a}.  The LBGs and BX objects show increasing
    clustering with increasing $R$-brightness, whereas the BM objects
    show no apparent trend, and the $K$-selected MUSYC galaxies appear
    to show the opposite trend.}
  \label{fig:rR}
\end{figure}

Next we split the sample into two populations using an apparent
optical magnitude cut $R = 25$, which is approximately equal to the
median total $R$-magnitude, and is $0.5$ magnitudes shallower than the
limit used for $z \sim 3$ LBGs \citep{steidel03}.  Fig.~\ref{fig:rR}
shows that the optically brighter $K$-selected subsample clusters less
strongly than the fainter subsample.  This result is at odds with
several studies of optically selected galaxies, including BX objects
at $z \sim 2.3$ \citep{adelberger05a}\footnote{It appears that for BX
  objects clustering strength increases with $R$-brightness,
  $K$-brightness, \emph{and} $R-K$ color \citep{adelberger05a,
    adelberger05b}; it follows that the faint blue BX objects are the
  least clustered.}, LBGs at $z \sim 3$ \citep{giavalisco01,
  foucaud03, adelberger05a, lee06}, and LBGs at $z \sim 4$ and $z \sim
5$ \citep{lee06, ouchi04}, all of which display stronger clustering
with increased rest-frame UV luminosity.  It is interesting that our
brighter subsample has a correlation length that agrees well with
R-selected samples in the same magnitude range.  Thus our results
suggest that the $K$-selected galaxies that are below the limits of
current $R$-selected surveys are the most strongly clustered.

We note that the median $K$-magnitudes of our two subsamples are
similar--with the optically-faint sample $0.1$ magnitudes brighter
than the optically-bright sample--and the overall distributions of
$K$-magnitude are also similar.  So the anti-correlation between $r_0$
and $R$-brightness may simply be a manifestation of the correlation
between $r_0$ and the $R-K$ color (Fig.~\ref{fig:rRK}).  Similarly,
the observations of \citet{shapley05} suggest a relationship between
$R-K$ and $K$ for BX objects, and \citet{adelberger05b} speculate that
the observed relationship between $r_0$ and $K$ for their sample may
reflect an underlying correlation between $r_0$ and $R-K$.  Thus the
results for both $K$-selected galaxies and optically-selected galaxies
suggest that color may be the most important driver of clustering
strength.  If this is the case, the difference in colors between these
two populations may explain the difference in clustering properties,
as $K$-selected galaxies tend to be redder.

\subsubsection{Clustering as a function of stellar mass}
\label{sec:stellarmass}

To investigate the relationship between stellar mass and clustering,
we estimate the mass of MUSYC galaxies by fitting \citet{bruzual03}
models to the observed photometry at fixed $z_{\rm{phot}}$.  We assume
a $\tau = 300 \textrm{Myr}$ declining star formation history, solar
metallicity, and a $0.1-100 M_\odot$ \citet{salpeter55} initial mass
function.

\begin{figure}
  \epsscale{1.1}
  \plotone{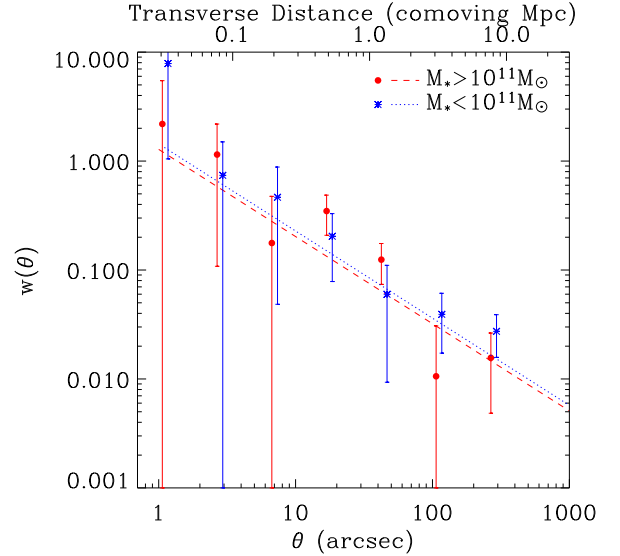}
  \caption{Same as Fig.~\ref{fig:wJK}, for mass-selected samples.}
  \label{fig:wmass}
\end{figure}

\begin{figure}
  \epsscale{1.1}
  \plotone{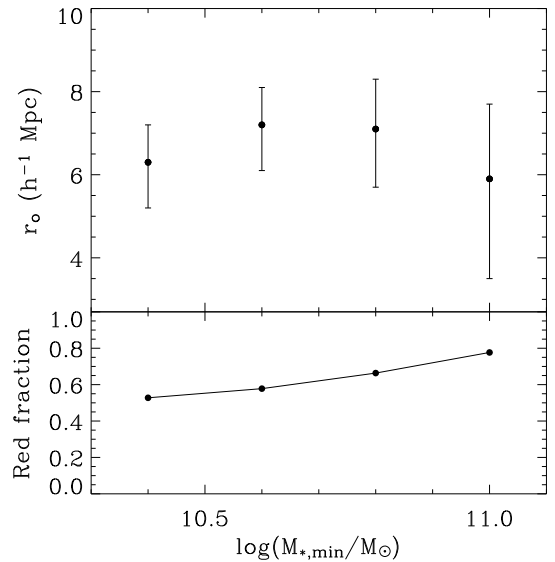}
  \caption{The comoving correlation length as a function of
    stellar mass threshold.  The lower plot shows the fraction of red
    galaxies (defined as galaxies redder than the median, $J-K \geq
    2.17$) as a function of mass.}
  \label{fig:rmass}
\end{figure}

Fig.~\ref{fig:wmass} shows that the angular correlation function of
galaxies with $M_* > 10^{11} M_\odot$ and for galaxies with $M_* <
10^{11} M_\odot$ are very similar.  The top panel of
Fig.~\ref{fig:rmass} shows the correlation length vs. minimum stellar
mass threshold.  There is no clear trend.  The bottom panel of
Fig.~\ref{fig:rmass} shows that the fraction of red galaxies is an
increasing function of mass threshold.  It was shown above that $r_0$
increases strongly with $J-K$ color; from this it might be expected
that clustering would increase strongly with mass, because the most
massive galaxies tend to be red.  Moreover, the median masses of the
$R$-faint and $R$-bright samples in Fig.~\ref{fig:rR} are $6.2 \times
10^{10} M_\odot$ and $1.6 \times 10^{11} M_\odot$, again suggesting a
possible relationship between mass and $r_0$.  So it is interesting
that we do not observe a clear relationship between clustering and
stellar mass, although it must be noted that the data shown in
Fig.~\ref{fig:rmass} have large error bars.

As discussed by \citet{vandokkum06}, the majority ($\sim 65\%$ in the
current sample) of $M_* > 10^{11} M_\odot$ galaxies at $z>2$ are DRGs.
With such significant overlap between the massive and red galaxies, it
is to be expected that they have similar correlation lengths.
However, we find $r_0 = 5.9^{+1.8}_{-2.4} h^{-1} \textrm{Mpc}$ for
$M_* > 10^{11} M_\odot$ galaxies, and $r_0 = 11.1^{+1.3}_{-1.4} h^{-1}
\textrm{Mpc}$ for DRGs.  This indicates that either the massive
non-DRGs have a very low correlation length, or that the less massive
DRGs have a very high correlation length.  Our sample is not large
enough to investigate each of these sub-populations individually, but
we do note that the correlation length for galaxies that are both
massive and meet the $J-K>2.3$ criterion for DRGs is $r_0 =
7.9^{+1.9}_{-2.4} h^{-1} \textrm{Mpc}$, intermediate between the
massive and DRG samples.  This may be evidence that the low-mass DRGs
are highly clustered \emph{and} that the high-mass blue galaxies are
less clustered.  Relative to the median high mass DRG, the median low
mass DRG is fainter in the NIR ($K=20.8$ vs. $20.3$), brighter in the
optical ($R=25.8$ vs. $27.1$), but has a similar NIR color ($J-K=2.6$
vs. $2.7$).  We have verified that it is not $K$-\emph{faint} DRGs
which contribute so strongly to the clustering but rather it is the
\emph{low-mass} DRGs by measuring $r_0 = 10.5^{+1.6}_{-1.8} h^{-1}
\textrm{Mpc}$ for $K<20.7$ DRGs.  This result is analogous to
conclusions from the local universe, where low-mass red galaxies and
high-mass red galaxies inhabit the densest environments
\citep{hogg03,kauffmann04}, and to the fact that, among massive
galaxies, there is a strong relationship between correlation length
and optical color \citep{li06}.  Additionally, \citet{kauffmann04}
show that the stellar mass of galaxies is not a strong function of
halo mass in the most massive halos.

It should be noted that, while we are approximately complete for
galaxies with $M_* > 10^{11} M_\odot$ \citep{vandokkum06}, we will be
very incomplete for less massive galaxies.  Using stellar mass
estimates from the ultradeep FIRES MS 1054-03 field \citep{forster06},
we estimate $\sim 65\%$ completeness for galaxies with $M_* >
10^{10.4} M_\odot$.  Scatter in our mass measurements may also obscure
any relationship between $r_0$ and mass.  Significantly deeper data
are needed to study the dependence of $r_0$ on mass for a complete
sample.

\section{Relationship between galaxies and dark matter halos}
\label{sec:halos}

In the halo model of galaxy formation, the galaxy correlation length
is related to the mass of dark matter halos \citep{mo96}.  In this
section we constrain the halo masses and occupation numbers of the
various subsamples of MUSYC galaxies using the measured correlation
lengths and number densities.

\subsection{The number density and bias of dark matter halos}

To investigate the relationship between the MUSYC $K$-selected
galaxies and dark matter halos, we use the halo mass function of
\citet{sheth99} that is derived from fits to large N-body simulations
\begin{equation}
\frac{dn_h}{dM} = A \left(1+\frac{1}{\nu'^{2q}}\right)\sqrt \frac{2}{\pi} \frac{\bar{\rho}}{M} \frac{d\nu'}{dM} \exp\left(\frac{-\nu'^2}{2}\right),
\label{mass_func}
\end{equation}
where $\nu' = \sqrt a \delta_c / \sigma(M,z)$, and the constants
$\delta_c \approx 1.69$, $a=0.707$, $A \approx 0.322$, $q=0.3$, and
$\bar{\rho}$ is the current mean mass density of the universe.  We
calculate the relative mass fluctuations in spheres that contain an
average mass $M$ as
\begin{equation}
\sigma(M,z) = D(z) \sigma(M,0),
\label{eq:sigma}
\end{equation}
where $D(z)$ is the growth factor for linear fluctuations given by
\citet{carroll92}, and $\sigma(M,0)$ is calculated using a scale-free
$n=1$ initial power spectrum and the transfer function of
\citet{bardeen86}.

The linear halo bias is calculated using the function of
\citet{sheth01}
\begin{equation}
b_h = 1 + \frac{1}{\delta_c} \left[ \nu'^2 + b\nu'^{2(1-c)} - \frac{v'^{2c}/\sqrt a}{\nu'^{2c}+b(1-c)(1-c/2)} \right],
\label{eq:bias_func}
\end{equation}
where $b=0.5$ and $c=0.6$.  Further details can be found in
e.g. \citet{mo02}.

Several definitions of the bias--which relates the clustering of
objects to that of the overall dark matter distribution-- in terms of
observable quantities appear in the literature.  We choose
\begin{equation}
b = \frac {\sigma_{8,gal}} {\sigma_{8}(z)},
\label{eq:gal_bias}
\end{equation}
where $\sigma_{8}(z)$ is the variance in $8 h^{-1} \textrm{Mpc}$
spheres and is calculated analogously to eq.~\ref{eq:sigma}.  If the galaxy
correlation function $\xi(r)$ is a power-law of the form
eq.~\ref{eq:xi} then it can be integrated to give the relative
variance
\begin{equation}
\sigma_{8,gal}^2 = \frac {72} {(3-\gamma)(4-\gamma)(6-\gamma)2^\gamma} \left( \frac {r_0}{8h^{-1} \rm{Mpc}} \right) ^\gamma
\label{eq:sigma8}
\end{equation}
\citep[][\S 36, \S 59]{peebles80}. 

We model the simple case of one galaxy per halo above a minimum halo
mass threshold, i.e.~a halo occupation number of 1.  More detailed
models, such as setting the halo occupation number equal to a power
law above some mass threshold \citep[e.g.][]{wechsler01} are beyond
the scope of this paper.  As we measure the bias of the galaxy samples
at scales larger than $\sim 1 \textrm{Mpc}$, we can associate the
observed bias with the linear bias of the host halos calculated with
eq.~\ref{eq:bias_func}, thereby providing an estimate of the halo
mass.

\begin{figure}
  \epsscale{1.1}
  \plotone{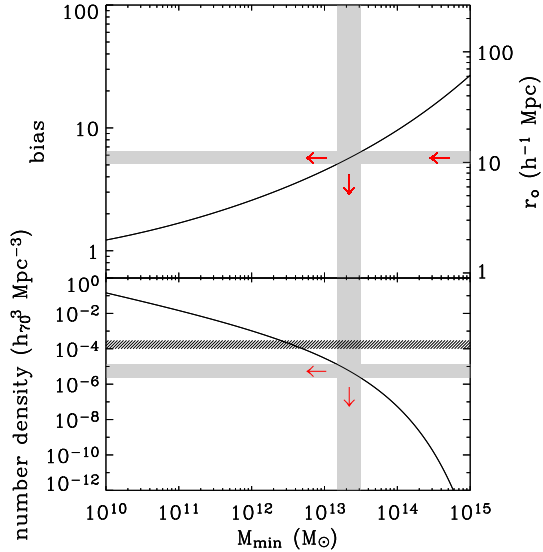}
  \caption{\emph{Upper:} The number-weighted average linear bias as a
    function of minimum halo mass threshold at $z=2.6$.  The right
    axis shows the relationship between galaxy correlation length and
    the inferred large-scale galaxy bias.  The shaded regions show the
    $1\sigma$ range of allowed $r_0$, and the corresponding range in
    bias and halo mass.  \emph{Lower:} The comoving halo number
    density as a function of halo mass threshold.  The shaded regions
    show the $1\sigma$ range of mass and number density for the halos
    that host DRGs.  The hatched region shows the observed number
    density of DRGs, illustrating that they are $40^{+60}_{-30}$ times
    more numerous than the halos that host them.}
  \label{fig:halomb_drg}
\end{figure}

Fig.~\ref{fig:halomb_drg} shows the average halo bias (weighted by the
number density) and number density as a function of halo mass
threshold at the median redshift, $z \simeq 2.6$.  The $1 \sigma$ bias
range for DRGs, as well as the minimum mass and the number density of
halos with the same bias values, is illustrated by the shaded region.
From this figure we can read off the mass threshold of halos that have
the same bias as DRGs, $M_h \approx 1.5-3 \times 10^{13} M_\odot$.
The larger sample of $K<21$ galaxies has $M_h \approx 1.5-5 \times
10^{12} M_\odot$.  For comparison, the LBGs and BX galaxies occupy
halos with mass threshold $M_h \approx 10^{11.5} M_\odot$ and $\approx
10^{12} M_\odot$, respectively \citep{adelberger05a} .

\subsection{The halo occupation number}

In the local universe, halos with mass $\lesssim 10^{12} M_\odot$ tend
to have only one bright galaxy, whereas $\gtrsim 10^{14} M_\odot$
halos may contain dozens \citep[e.g.][]{kauffmann04}.  There is also
evidence of galaxies sharing halos at high redshift
\citep{daddi03,zheng04,adelberger05a,ouchi05,lee06}.  We define the
mean halo occupation number $N_{occup}$ as the ratio of galaxy number
density to the number density of host dark matter halos.  Occupation
numbers greater than unity suggest that multiple galaxies can reside
in a single halo.

The simplest estimate for the number density of galaxies comes from
dividing the observed number of galaxies by the volume probed by our
survey at $2<z<3.5$.  However there may be significant evolution of
the actual number density over this redshift range, and there is
probable contamination by interlopers.  We attempt to correct for
these effects by estimating the fraction of the observed galaxies that
lie at $z \sim 2.6$ using the redshift distributions discussed in \S
\ref{sec:data}.  We note that these two estimates agree to within the
field-to-field variance within the survey, which is $\sim 20\%$.  The
estimated number density of galaxies at $z \sim 2.6$ with $K<21$ is
$(5 \pm 2.5) \times 10^{-4} h_{70}^3\textrm{Mpc}^{-3}$.  For DRGs, we
estimate $(2 \pm 1) \times 10^{-4} h_{70}^3\textrm{Mpc}^{-3}$.
GOODS-CDFS -- the only other public field with size, depth, and
multiwavelength coverage comparable to one of our fields -- shows a
lower density of massive and red galaxies than are present in the
MUSYC survey, indicating that our field-to-field variance may not be
representative \citep{vandokkum06, grazian06}.  Incompleteness and
possible systematics in photometric redshifts further complicate
density estimates; we therefore assign approximate $50\%$
uncertainties.  These number densities are consistent with estimates
from the luminosity function \citep{marchesini06}.

\begin{figure}
  \epsscale{1}
  \plotone{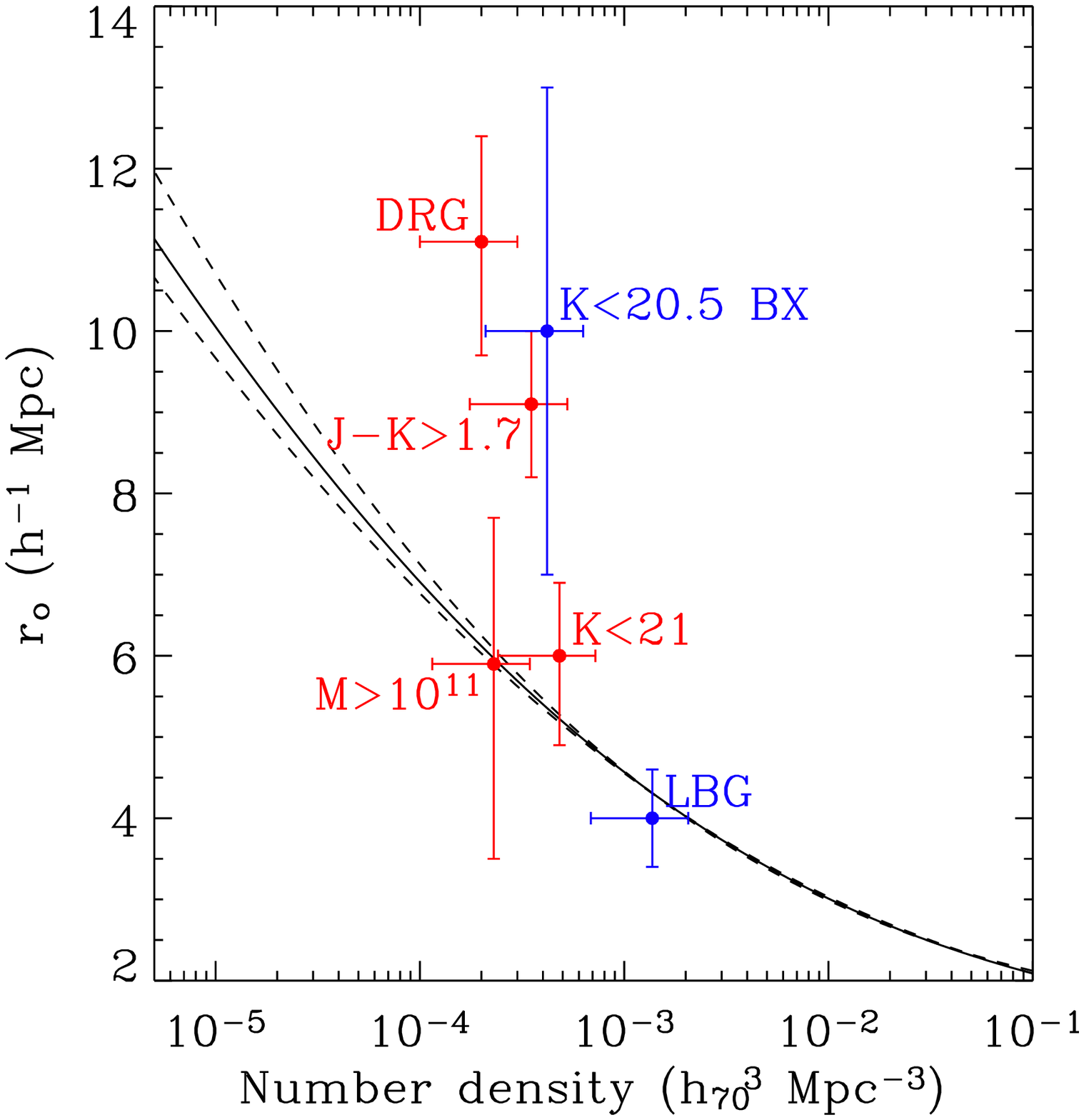}
  \caption{The correlation length and the number density of different
    populations of $z > 2$ galaxies.  The LBG data are taken from
    \citet{adelberger05a}.  The $r_0$ for the $K$-bright BX galaxies
    is taken from \citet{adelberger05b}.  We estimate the number
    density of bright BX galaxies by using density of all BX galaxies
    given by \citet{adelberger05a} and applying a correction using the
    information given in \citet{adelberger05b}.  The solid line shows
    the approximate correlation length $r_0$ as a function of number
    density that would be expected in the case of one galaxy per halo
    at $z=2.6$; the upper and lower dashed lines show the same
    information at $z=3$ and $z=2$, respectively.  Although the
    inferred halo mass for a galaxy population with a given $r_0$ will
    depend on the redshift of the galaxy population, the relationship
    between $r_0$ and number density for dark matter halos is not a
    sensitive function of redshift.}
  \label{fig:ro_n}
\end{figure}

Fig.~\ref{fig:ro_n} compares the number density and correlation
lengths for various samples to the values that are expected in the
case of a one-to-one relationship between galaxies and dark matter
halos.  As noted by e.g. \citet{adelberger05a}, the observed
properties of $z \sim 3$ LBGs are roughly consistent with such a
relationship.  The same is true for the entire MUSYC $K$-selected
sample, as well as the sample of $M_* > 10^{11} M_\odot$ galaxies.
However the redder galaxies deviate strongly from the expected
relation, suggesting high occupation numbers.  In particular, the
number density of halos with the same bias as DRGs is $\approx 2.4-13
\times 10^{-6} h_{70}^3\textrm{Mpc}^{-3}$, suggesting $N_{occup}
\approx 40^{+60}_{-30}$.  However, if the estimated field to field
variance is taken into account (Table \ref{tbl:results}), the DRGs may
have a correlation length as low as $r_0 \approx 7 h^{-1}
\textrm{Mpc}$, in which case $N_{occup} \sim 1$. The bright $K<20.5$
BX galaxies from \citet{adelberger05b} also suggest high occupation
numbers, although the very strong clustering in one of their observed
fields may drive their result.

\subsection{Constraining the occupation number with the number of
  close pairs}

In the previous subsection we estimated the occupation number
$N_{occup}$ using the standard method of comparing the observed number
density of galaxies to the number density of their host halos.  If
many galaxies share the same halo, then the angular positions of
galaxies should show strong `clumps,' i.e. the correlation function
should indicate large power on small scales.  Here we obtain an
independent estimate of $N_{occup}$ by comparing the shape of the
angular correlation function $w(\theta)$ to the angular correlation
function of dark matter halos $w_h(\theta)$.

Although the angular and spatial correlation functions of dark matter
halos are well-approximated by power laws on scales larger than $\sim
r_{vir}$, the probability of finding \emph{distinct} halos with
separation $<r_{vir}$ necessarily falls to $0$; this sets
$\xi_h(r<r_{vir}) \sim -1$.  However, for a sufficiently large
redshift selection window, $w_h(\theta)$ will tend to remain flat on
small angular scales because of projection effects (see
Fig.~\ref{fig:wK21}).  The extent to which the galaxy correlation
function $w(\theta)$ differs from the $w_h(\theta)$ on small scales
can be used to constrain the occupation number \citep{wechsler01,
  bullock02, adelberger05a}.

The expected number of galaxies in the angular interval
$(0,\theta_{max})$ around a randomly-chosen galaxy is
\begin{equation}
N = n_g \pi \theta_{max}^2 (1 + w_{<\theta_{max}}),
\label{eq:expected_pairs}
\end{equation}
where $n_g$ is the mean surface density of galaxies, and
$w_{<\theta_{max}}$ is the value of the angular correlation function
evaluated over the same angular interval (i.e. using
eq.~\ref{eq:landyszalay} and a bin of width $\theta_{max}$).  This
relation follows directly from the definition of the angular
correlation function.  If all galaxies are associated with halos, and
if $\theta_{max}$ is larger than the virial radius, the average number
of galaxies within the same angular interval is the number of
additional galaxies within the host halo plus a contribution from
neighboring halos,
\begin{equation}
N \approx f + N_{occup} n_h \pi \theta_{max}^2 (1 + w_{h,<\theta_{max}}),
\label{eq:expected_pairs_halomodel}
\end{equation}
where $f$ denotes the average number of additional galaxies in a halo
that hosts at least one galaxy.  Note that $N_{occup}=n_g/n_h$, so the
right hand side of eq.~\ref{eq:expected_pairs_halomodel} does not
depend directly on $N_{occup}$.  Combining
eqs.~\ref{eq:expected_pairs} and \ref{eq:expected_pairs_halomodel}
gives an estimate of $f$ that depends on both $w(\theta)$ and
$w_h(\theta)$ over the interval $(0,\theta_{max})$, as well as on
$n_g$.  Under the assumption that all halos above the minimum mass
threshold host detectable galaxies, it is apparent that $N_{occup} =
(f+1)$.  If some fraction $g$ of these halos do not host detectable
galaxies, then $N_{occup}$ is reduced by a factor $(1-g)$.  There is
some indication that $g > 0$ for LBGs \citep{adelberger05a, lee06}.
Here we make the simplifying assumption that $g=0$, which is
consistent with the $N_{occup} \gtrsim 1$ measurements from the
previous subsection, and note that the $N_{occup}$ measurements of
this section may be upper limits.

We measure $w_{h,<\theta_{max}}$ directly from the observing cone output of
the GalICS simulations (see \S \ref{sec:simulations}), using halos
with a large-scale angular correlation and redshift distribution
similar to those inferred for our sample.  Because the GalICS observing
cone outputs do not specify the coordinates of halos, we instead use
the coordinates of the most central galaxy in each halo when measuring
the correlation functions.  As shown in Fig.~\ref{fig:wK21},
$w_h(\theta)$ flattens over the region $\theta \lesssim 40\arcsec$.
Our results are relatively insensitive to the details of this
procedure, and the uncertainties are dominated by the uncertainties in
$w(\theta)$ and in the galaxy surface density $n_g$, which we estimate
from the observed variance among the MUSYC fields.  We use the
observed field-to-field variations to estimate the uncertainty in
$n_g$, and use $\theta_{max}=60\arcsec$ in
eqs.~\ref{eq:expected_pairs} and \ref{eq:expected_pairs_halomodel}.

We derive $N_{occup} \approx 1.7 \pm 0.3$ for DRGs.  Other galaxy
subsamples have occupation numbers that are consistent with the DRG
value to within $1\sigma$.  In the previous subsection we showed that
the observed correlation lengths and number densities suggest $N_{occup}
\sim 1$ for the $K<21$ and $M_* \sim ~10^{11} M_\odot$ galaxies,
consistent with the values derived here.  However, the value for DRGs is
much less than the $N_{occup} = 40^{+60}_{-30}$ that is inferred from
the correlation length.  This may indicate that our measured
correlation length is an overestimate.  Possible causes for this
discrepancy are given in \S \ref{sec:discussion}, however we note that
these values are consistent if the estimated field-to-field variance
in the correlation length is taken into account (\S
\ref{sec:simulations} and Table \ref{tbl:results}).  Previous studies
have found correlation lengths even larger than ours
\citep{daddi03,grazian06}, although their measurements may have been
unduly influenced by small-scale structure in $w(\theta)$ \citep[\S
\ref{sec:clustresults}, and][]{zheng04}.

In the previous subsection we showed that the number density and
correlation length of $K<20.5$ BX galaxies \citep{adelberger05b} point
toward very high occupation numbers.  However these authors indicate
that their results do not change significantly if they include galaxy
pairs at small separations in their analysis.  This suggests that
there is no evidence of a strong small-scale excess in their
correlation functions, and therefore that bright BX galaxies also show
a disagreement between their observed properties and the properties of
dark matter halos, although the difference is not as significant as
for DRGs.

\section{Relating galaxy populations at different redshifts}

\subsection{Evolution of galaxy bias}

What are the $z=0$ descendants of the $K$-selected galaxies discussed in
this paper?  In the $\Lambda\textrm{CDM}$ picture of structure
formation, the large-scale distribution of galaxies is determined
primarily by the dark matter potential wells.  Thus we can address the
evolution of high-redshift galaxies by following the evolution of
their host dark matter halos.  Fortunately, the dynamics of
collisionless dark matter particles are well-described by simple
models or by cosmological N-body simulations.  So while the
complicated physics that dictates the evolution of the baryonic
component of galaxies (e.g.~star formation, feedback, mergers) cannot
be addressed by these models or simulations alone, we can constrain
the bias and halo mass of the $z=0$ descendants.

\begin{figure}
  \epsscale{1.1}
  \plotone{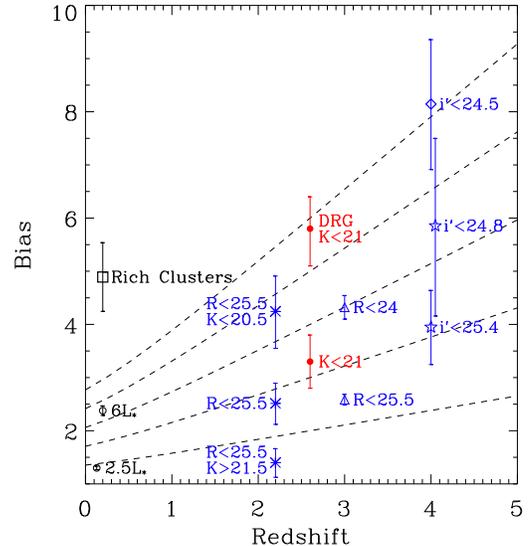}
  \caption{The evolution of bias with redshift.  The tracks show the
    evolution of bias calculated using the galaxy-conserving model,
    eq.~\ref{eq:bias_ev}.  The filled red circles are based on this
    work, while the blue symbols are for optically-selected galaxies
    at various redshifts and black symbols are for local galaxies.
    The asterisks show the bias of BX galaxies from
    \citet{adelberger05a,adelberger05b}.  The triangles show the bias
    of $z \sim 3$ LBGs from \citet{lee06}.  The diamond and stars are
    for the $z \sim 4$ LBGs of \citet{allen05} and \citet{ouchi04},
    respectively.  Open circles are from \citet{zehavi05}.  The square
    represents the richest cluster sample analyzed by
    \citet{bahcall03}.  In all cases, we have calculated the bias
    using eqs.~\ref{eq:gal_bias} and \ref{eq:sigma8}.}
  \label{fig:bias_ev}
\end{figure}

One way to investigate relationships between galaxy populations at
different redshifts is to compare their observed bias.  As the
universe evolves with time, the dark matter becomes more clustered and
the bias of a set of biased objects will decrease.  The bias $b(z)$
of a set of test particles evolved according to
\begin{equation}
b(z) = 1+(b(0)-1)/D(z)
\label{eq:bias_ev}
\end{equation}
\citep{fry96} where $D(z)$ is the growth factor.  It is important to
note that this equation does not account for merging and the evolution
of the baryonic components of galaxies.  Merging will play a role if
galaxies in the densest (i.e.~most biased) regions of space are more
likely to merge than galaxies in less dense regions, thereby reducing
the average bias of unique descendants.  So it is possible that the
bias will evolve faster than indicated by eq.~\ref{eq:bias_ev}.  We
refer to this as the `galaxy-conserving' model of bias evolution.
Fig.~\ref{fig:bias_ev} shows tracks of bias evolution, along with the
bias of different samples of galaxies (which we compute in a
consistent way, using eq.~\ref{eq:gal_bias}).  This figure shows that
the brightest LBGs at $z \sim 3$ \citep[$R<24$;][]{lee06} have a bias
roughly consistent with the $z \sim 2.6$ $K<21$ galaxies studied here.
At higher redshift, only the brightest $z \sim 4$ LBGs
\citep[$i'<24.8$;][]{ouchi04,allen05} show biasing consistent with the
lower-redshift DRGs, but the fainter $z \sim 4$ LBGs are not
consistent.

Fig.~\ref{fig:bias_ev} shows that all of the high-redshift samples
discussed in this paper, including LBGs, evolve into highly-biased
populations at $z \sim 0$.  This point has been made previously for
the optically-selected populations
\citep[e.g.][]{baugh98,ouchi04,adelberger05a}.  Among the known
galaxies at $z \sim 2$, it appears that only those that are faintest
in $K$ can be progenitors of typical $L_*$ field galaxies.

\subsection{Evolution of halo mass}

The preceding analysis only illustrates the bias of the descendants of
high-redshift galaxies.  If the galaxies within a given population
follow different evolutionary paths, then the descendants will have
diverse properties, and the average bias will have limited
interpretive value.  Knowledge of the range of environments or halo
masses of the descendants is more meaningful.  For a more detailed
investigation of the $z=0$ descendants of $z>2$ galaxies, we track the
growth of dark matter halos using cosmological N-body simulations.  We
choose the GIF simulation for its size and mass resolution, and for
the publicly available halo catalogs and merger trees \citep{frenk00}.
This simulation uses $\Omega_m = 0.3$, $\Omega_\Lambda = 0.7$, $h =
0.7$, $\Gamma = 0.21$, and $\sigma_8 = 0.9$.  The linear size is
$141.3 h^{-1}$ comoving Mpc.  A minimally-resolved halo consists of 10
particles with mass $1.4 \times 10^{10} h^{-1} M_\odot$, but we note
that the merging histories of halos less massive than $\sim 100$
particles may be inaccurate \citep{kauffmann99}.

In these simulations, the position of a central galaxy within a halo
is given by the position of the most-bound dark matter particle.  When
halos merge, the central galaxy of the most massive progenitor becomes
the new central galaxy while the central galaxy of less massive
progenitors, as well as any progenitor satellites, are kept as
satellites in the descendant halo.  We assume that the galaxies in the
present sample begin as central galaxies in $z \sim 2.6$ halos more
massive than the threshold masses given in the previous subsection,
and follow the positions of these galaxies to $z = 0$.  Occasionally
satellite galaxies are ejected during mergers and may not be contained
in any of the simulated halos at later times, however this effect
occurs rarely for the massive halos considered here, and may be safely
ignored.  Additionally, we exclude halos near the edges of the
simulation from analysis.

We note that our treatment of halo evolution is different than that of
many other authors, and these differences may lead to contrasting
conclusions.  For instance, \citet{grazian06} employ the "merging
model" of \citet{matarrese97} and \citet{moscardini98} as one method
to study the clustering evolution of DRGs.  In the context of this
model, $z=0$ galaxies are considered to be descendants of DRGs if they
occupy halos that are more massive than the $z>2$ DRG host halos.
More halos will meet this mass threshold at $z=0$ than at $z>2$, so
many of these descendants enter the sample at intermediate redshifts,
leading to a lower typical descendant halo mass.  In this section we
are concerned only with the \emph{direct} descendants of $z \sim 2.6$
MUSYC galaxies.

\begin{figure}
  \epsscale{1.1}
  \plotone{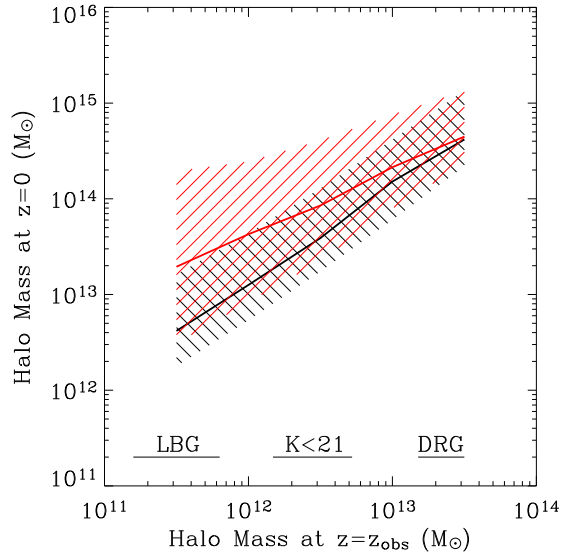}
  \caption{The evolution of halo mass with redshift.  The x-axis shows
    the halo mass at the redshift of observation, and the y-axis shows
    the mass of the descendant halos at z=0.  The red hatched region
    indicates the predicted $68\%$ halo mass range for the descendants
    of $z \sim 2.6$ galaxies under the assumption of no galaxy
    mergers.  The black hatched region indicates the mass range under
    the assumption that all galaxies that share halos merge.  See the
    text for details.  The thick solid curves show the median halo
    masses under each of these scenarios.  The horizontal lines at the
    bottom of the figure mark the range of halo masses inferred for
    LBGs \citep{adelberger05a}, $K<21$ galaxies, and DRGs at the epoch
    of observation.}
  \label{fig:halo_ev}
\end{figure}

There is an essential ambiguity in interpreting the range of halo
masses occupied by the descendants of high-redshift galaxies.  The
simulations show that--because of halo mergers--many $z=0$ halos host
multiple descendants.  However it is unclear whether the galaxies
themselves merge or whether they retain separate identities within a
single halo.  First we deal with the scenario in which the galaxies do
not merge, as in the galaxy-conserving model discussed above.  The red
hatched region in Fig.~\ref{fig:halo_ev} shows the $68\%$ range of
host halo masses for descendants of high-redshift galaxies, as a
function of halo mass at $z \sim 2.6$.  The descendants of DRGs
primarily occupy cluster-scale halos, with mass $\gtrsim 10^{14}
M_\odot$.  If our estimate of the correlation lengths for DRGs is
correct, then DRGs with $K<21$ cannot be progenitors of the majority
of local field early-type galaxies.  It follows that DRGs exist in
proto-cluster regions.  The majority of LBGs also end up in group and
cluster-scale halos.  The black hatched region in
Fig.~\ref{fig:halo_ev} shows the mass range of halos occupied by
descendants under the assumption that all galaxies within a single
halo merge.  In this case the most massive halos at $z=0$, which host
multiple descendants of high-redshift galaxies, are only counted once.
The difference between the two hatched regions illustrates the
importance of merging; for instance, $\gtrsim 50\%$ of the halos that
are inferred to host LBGs merge with a more massive halo between $z
\sim 3$ and $z \sim 0$.  This indicates either that LBGs tend to merge
with more massive galaxies to form the brightest central galaxy in a
halo, or that LBG descendants are satellites rather than the brightest
central galaxy in these halos.  It may be possible that some of the
progenitors of the low-mass red galaxies that exist in dense regions
\citep[e.g.][]{hogg03} are LBGs, although this would require star
formation to cease shortly after the epoch of observation.

\section{Summary and Discussion}
\label{sec:discussion}

We have used $300 \textrm{arcmin}^2$ of $UBVRIzJHK$ imaging from the
MUSYC survey to study the angular and spatial correlation functions of
$K$-selected galaxies with $2<z_{\rm{phot}}<3.5$ and $K<21$.  The
correlation length for this sample is $r_0 = 6.0^{+0.9}_{-1.1} h^{-1}
\textrm{Mpc}$, $\sim 50\%$ larger than optically-selected galaxies at
similar redshifts \citep{adelberger05a, lee06}.  The clustering of
galaxies increases strongly with $J-K$ and $R-K$ color, and the
$J-K>2.3$ population of distant red galaxies has $r_0 =
11.1^{+1.3}_{-1.4} h^{-1} \textrm{Mpc}$.  Our results for DRGs are
lower than previous results \citep{daddi03,grazian06}.  This may be
partially due to our smaller uncertainties.  Additionally, previous
studies were only able to constrain the correlation function on small
scales, where the signal may be strongly affected by galaxies that
share the same halo \citep{zheng04, ouchi05, lee06}.  In contrast, we
perform fits to the correlation function on scales larger than the
typical halo virial radius where this effect is reduced.

Nevertheless we confirm the basic trend indicated by previous studies,
in which red galaxies are much more strongly clustered than
optically-selected galaxies.  Moreover, this trend is not simply due
to the fact that the red galaxies used in the previous studies were
selected in the $K$-band: clustering increases strongly with color
even within our $K$-selected sample, while there is no significant
relationship between $K$-magnitude and color or between $K$-magnitude
and correlation length.  These results suggest that a color-density
relationship was in place at $z>2$, as is also indicated by the
properties of optically-selected galaxies in different environments
\citep{steidel05}.  Whether this relationship is driven by galaxies
that are red because of dust obscuration, or because of low specific
star formation rates, remains to be seen.

\citet{cucciati06} and \citet{cooper06} investigate the fraction of
red galaxies as a function of local galaxy density, finding that the
color-density relation extends only to $z \sim 1.3-1.5$.  This
apparent contradiction with our results may be due to the difference
in measurement techniques (fraction of red galaxies vs. correlation
lengths).  These studies may also be dominated by galaxies that are
less massive than the $z \sim 1.5$ descendants of MUSYC galaxies, so
there may not be a contradiction if the evolution of the color-density
relationship is mass-dependent \citep{cooper06}.  Finally, these
studies are based on spectroscopy of galaxies that are selected in the
optical, leaving the possibility that they are incomplete for the
reddest galaxies at these redshifts.

A color-density relation at $z>2$ has implications for current galaxy
selection techniques.  \citet{adelberger05b} show that, among an
optically-selected sample of $z \sim 2.2$ BX objects, there is a
strong correlation between clustering strength and $K$-band magnitude.
Additionally, \citet{shapley04} show that these galaxies are also
massive, metal rich, and have high star formation rates.  Taken
together, these results indicate that the $K$-bright galaxies found by
optical surveys show similar properties to those uncovered by NIR
surveys \citep[e.g.][]{franx03, daddi03, daddi04, vandokkum04,
  vandokkum06, kriek06a}.  It has been argued that the primary
difference between optically-bright massive galaxies and
optically-faint massive galaxies may be that the former are observed
during a chance period of unobscured star formation
\citep[e.g.][]{shapley05}.  However, the observed relationship between
clustering and $J-K$ or $R-K$ color suggest that this is not the
complete explanation.  For instance, among our $K$-selected sample, it
is the \emph{R-faint} galaxies \citep[i.e.~those that could not make
it into the sample of][]{adelberger05b} that cluster most strongly.
Additionally, the very high correlation lengths measured by
\citet{adelberger05b}, $r_0 \sim 10 h^{-1} \textrm{Mpc}$ for galaxies
with $K \sim 20.5$, may not be representative because one of their
four fields shows anomalously high clustering; the average of the
other three fields is $r_0 = 5 \pm 1 h^{-1} \textrm{Mpc}$ (see their
Fig.~2).  Applying the same $K < 20.5$, $R < 25.5$ selection criteria
to the MUSYC sample, we find $r_0 = 4.5^{+2.7}_{-4.5}
h^{-1}\textrm{Mpc}$.  We conclude that the difference between massive
optically-selected galaxies and massive red $K$-selected galaxies is
probably not ``transient'', but is instead related to a fundamental
difference in the host dark matter halos.

It is interesting to consider whether $r_0$ is primarily related to
stellar mass, or some other property.  We use stellar population
models to estimate the stellar masses, and do not find a significant
relationship between mass and $r_0$.  This is interesting, as there is
a strong relationship between color and mass; together, these results
indicate that either low-mass red galaxies cluster very strongly or
that massive blue galaxies cluster less strongly.  We find some
evidence supporting both of these conclusions.  We discuss our
findings in light of the recent results from optically-selected
samples, finding that there is evidence that redder colors are
associated with larger correlation lengths for both samples.  For
optically-selected galaxies, the observed relationship between
decreasing $R$-magnitude and increasing clustering strength for LBGs
has been taken to suggest a positive relationship between halo mass
and star formation rate at early times \citep[e.g.][]{giavalisco01}.
On the other hand, stellar population synthesis modelling has not
uncovered a relationship between rest-frame UV luminosity and stellar
mass for typical optically-selected galaxies \citep[][but see
\citealt{papovich01} for a discussion of a fainter sample]{shapley01,
  shapley05}.  Similarly, \citet{adelberger05b} show that the
clustering of optically-selected BX galaxies is strongly related to
$K$-brightness.  While there is a correlation between $K$ and stellar
mass \citep{shapley05}, there is also a relationship between $K$ and
star formation rate \citep{reddy05,erb06}.  So while firm conclusions
cannot be drawn from this evidence, it appears that clustering may not
be determined by stellar mass alone for optically-selected samples.
The evidence presented here for $K$-selected galaxies also indicates
that color, rather than stellar mass, may be the primary determinant
of $r_0$.

We compared the observed clustering length of MUSYC galaxies to that
of dark matter halos, assuming the simple case of one galaxy per halo
above a halo mass threshold.  Galaxies with $2<z_{\rm{phot}}<3.5$ and
$K<21$ occupy halos with $M \gtrsim 3 \times 10^{12} M_\odot$.  The
number density of galaxies is similar to the number density of halos,
indicating a mean halo occupation number of order $\sim 1$.  Galaxies
with redder colors reside in more massive halos, with DRGs residing in
$M \gtrsim 10^{13} M_\odot$ halos.  However DRGs are found to be more
numerous than these halos, suggesting a mean halo occupation number
$N_{occup} \approx 40^{+60}_{-30}$.  If such large numbers of galaxies
occupy the same halos, then it is expected that the angular
correlation function $w(\theta)$ will show very large values on scales
corresponding approximately to the halo virial radius.  We performed
an independent estimate of the occupation numbers by comparing the
small-scale values of $w(\theta)$ to the values expected for dark
matter halos; this yields occupation numbers $\sim 1.5-2$ for all
samples studied in this paper, regardless of color and stellar mass.

The cause of the discrepancy in the occupation number of red galaxies
is unclear.  One possibility is that we have overestimated the
correlation length of DRGs, although previous studies
\citep{daddi03,grazian06} have found correlation lengths even larger
than the value presented here.  It is important to note that the high
$r_0$ measured for DRGs is largely due to the integral constraint
correction (\S \ref{sec:method}); neglecting this correction decreases
the power law amplitude of $w(\theta)$ by $\sim 50\%$, and decreases
the correlation length from $r_0 = 11.1^{+1.3}_{-1.4} h^{-1}
\textrm{Mpc}$ to $r_0 = 7.7^{+1.6}_{-1.9} h^{-1} \textrm{Mpc}$.
Larger fields are necessary to reduce the effect of the integral
constraint.  We also note that occupation numbers greater than unity
indicate that more complex halo occupation distribution models--in
which the number of galaxies per halo depends on the halo mass--are
necessary in order to accurately quantify the relationship between
galaxies and halos \citep[e.g.][]{zheng04,lee06}.  Regardless of the
cause of this discrepancy, the relationship between color and
clustering in our sample has been established with $\sim 96\%$
significance.

Finally, we addressed the evolution of $z>2$ galaxies.  The
descendants of our $K$-selected populations tend to occupy halos with
masses $10^{13} - 10^{14} M_\odot$, corresponding to the mass scales
of groups.  The reddest samples, including the DRGs, may occupy
cluster-scale halos, with masses $\gtrsim 10^{14} M_\odot$.  Even the
descendants of the less clustered LBGs tend to reside in groups.  It
appears that only a small subset of the $z>2$ galaxies that dominate
current redshift surveys can be progenitors of typical $L_*$ field
galaxies.

An important caveat is that each galaxy `population' will have rather
heterogeneous properties, and it may be that discussing the
evolutionary paths of population averages obscures important
distinctions.  For instance, \citet{adelberger05a} show that the
correlation length of BX objects is $\sim 4 h^{-1} \textrm{Mpc}$, and
argue from this that their descendants should be elliptical galaxies.
\citet{adelberger05b} show that the BX objects with $K \lesssim 21$
contribute most strongly to the clustering measurement; $40\%$ of the
BX objects are at $K>21.5$, and have a correlation length $\sim 2.5
h^{-1} \textrm{Mpc}$.  Presumably these fainter galaxies will evolve
into a much less clustered population by $z=0$.

A principle limitation of the preceding analysis is the heavy reliance
on photometric redshifts.  Obtaining large numbers of spectroscopic
redshifts for red $K$-selected galaxies has proven difficult on 6-10m
telescopes.  While NIR spectroscopy yields a high success rate for
determining redshift for bright galaxies \citep{kriek06b}, the advent
of multi-objects NIR spectroscopy will make the process more
efficient.  Another limitation of our study is the small galaxy
sample.  Recent clustering measurement of optically-selected galaxies
are based on samples that are 1-2 orders of magnitude larger than the
one presented here.  The next generation of NIR detectors will enable
the imaging of significantly larger fields to comparable depth,
allowing for more precise measurements of galaxy clustering.

\acknowledgements

We thank the members of the MUSYC collaboration for their contribution
to this research.  Feedback from the anonymous referee helped to
improve this paper.  MUSYC has greatly benefited from the support of
Fundaci\'{o}n Andes.  We are also grateful to the Lorentz Center
(Universiteit Leiden) for providing a venue for constructive meetings.
PGvD acknowledges support from NSF CAREER AST-0449678. DM is supported
by NASA LTSA NNG04GE12G. EG is supported by NSF Fellowship
AST-0201667.  PL is supported by Fondecyt Grant \#1040719.

Facilites: \facility{Blanco(ISPI)},\facility{Blanco(MOSAIC II)}

\clearpage

\begin{deluxetable}{ccccc}
\tablecolumns{5}
\tablecaption{Galaxy correlation functions: fitting range $0\arcsec<\theta<200\arcsec$}
\tablehead{ \colhead{Selection \tablenotemark{a}} & \colhead{$N_{gal}$} 
& \colhead{$A_w (\beta=0.8)$} & \colhead{$r_0 (\gamma=1.8)$} & \colhead{$r_0 (\gamma=1.6)$} }
\startdata

$K<21$ & $644$ & $1.9 \pm 0.3$ & $7.6^{+0.6}_{-0.6}$ & $9.1^{+0.7}_{-0.8}$ \\

& & & & \\

$J-K>1.1$ & $638$ & $1.8 \pm 0.3$ & $7.4^{+0.6}_{-0.7}$ & $8.9^{+0.7}_{-0.8}$ \\
$J-K>1.4$ & $614$ & $1.8 \pm 0.3$ & $7.4^{+0.6}_{-0.7}$ & $8.9^{+0.8}_{-0.8}$ \\
$J-K>1.7$ & $493$ & $2.3 \pm 0.4$ & $8.3^{+0.7}_{-0.7}$ & $10.6^{+0.8}_{-0.9}$ \\
$J-K>2.0$ & $381$ & $2.8 \pm 0.5$ & $9.2^{+0.8}_{-0.9}$ & $12.1^{+1.0}_{-1.0}$ \\
$J-K>2.3$ & $267$ & $4.9 \pm 0.7$ & $12.0^{+0.9}_{-1.0}$ & $16.2^{+1.2}_{-1.2}$ \\

& & & & \\

$R-K>2.9$ & $626$ & $1.9 \pm 0.3$ & $7.5^{+0.6}_{-0.7}$ & $9.1^{+0.7}_{-0.8}$ \\
$R-K>3.4$ & $563$ & $2.1 \pm 0.3$ & $8.1^{+0.6}_{-0.7}$ & $10.2^{+0.8}_{-0.8}$ \\
$R-K>3.9$ & $444$ & $2.9 \pm 0.4$ & $9.6^{+0.7}_{-0.8}$ & $12.5^{+0.9}_{-0.9}$ \\
$R-K>4.4$ & $353$ & $3.9 \pm 0.5$ & $11.1^{+0.9}_{-0.9}$ & $15.0^{+1.1}_{-1.1}$ \\

& & & & \\

$R<25$ & $341$ & $1.9 \pm 0.5$ & $7.3^{+1.0}_{-1.2}$ & $8.7^{+1.3}_{-1.4}$ \\
$R>25$ & $303$ & $3.4 \pm 0.6$ & $10.4^{+1.0}_{-1.1}$ & $13.5^{+0.5}_{-1.4}$ \\

& & & & \\

$K>19.3$ & $620$ & $2.1 \pm 0.3$ & $7.9^{+0.6}_{-0.6}$ & $9.6^{+0.7}_{-0.8}$ \\
$K>19.7$ & $574$ & $2.4 \pm 0.3$ & $8.6^{+0.6}_{-0.7}$ & $10.4^{+0.8}_{-0.8}$ \\
$K>20.1$ & $480$ & $2.6 \pm 0.4$ & $8.9^{+0.7}_{-0.7}$ & $10.8^{+0.9}_{-0.9}$ \\
$K>20.5$ & $279$ & $2.7 \pm 0.7$ & $8.9^{+1.2}_{-1.3}$ & $11.5^{+1.4}_{-1.6}$ \\

& & & & \\

$log(M)>10.4$ & $616$ & $2.1 \pm 0.3$ & $8.0^{+0.6}_{-0.6}$ & $9.7^{+0.7}_{-0.8}$ \\
$log(M)>10.6$ & $543$ & $2.0 \pm 0.3$ & $7.6^{+0.7}_{-0.7}$ & $9.4^{+0.8}_{-0.9}$  \\
$log(M)>10.8$ & $429$ & $1.8 \pm 0.4$ & $7.4^{+0.9}_{-1.0}$ & $9.3^{+1.1}_{-1.2}$  \\
$log(M)>11.0$ & $325$ & $2.1 \pm 0.6$ & $8.0^{+1.1}_{-1.3}$ & $9.6^{+1.4}_{-1.5}$  \\

\enddata
\tablenotetext{a}{All galaxies are selected using $K < 21$ unless otherwise specified}
\label{tbl:results_allrad}
\end{deluxetable}

\begin{deluxetable}{ccccccc}
\tablecolumns{7}
\tablecaption{Galaxy correlation functions and bias: fitting range $40\arcsec<\theta<200\arcsec$}
\tablehead{ \colhead{Selection \tablenotemark{a}} & \colhead{$N_{gal}$} 
& \colhead{$A_w (\beta=0.8)$} & \colhead{$r_0 (\gamma=1.8)$} & \colhead{$r_0 (\gamma=1.6)$} 
& \colhead {bias} & \colhead{total variance\tablenotemark{b}}}
\startdata

$K<21$ & $644$ & $1.3 \pm 0.4$ & $6.0^{+0.9}_{-1.1}$ & $6.6^{+1.0}_{-1.1}$ & $3.3^{+0.5}_{-0.5}$ & $^{+2.1}_{-3.0}$ \\

& & & & & & \\

$J-K>1.1$ & $638$ & $1.3 \pm 0.4$ & $6.1^{+0.9}_{-1.0}$ & $6.7^{+1.0}_{-1.1}$ & $3.4^{+0.5}_{-0.5}$ & $^{+2.0}_{-3.0}$ \\
$J-K>1.4$ & $614$ & $1.4 \pm 0.4$ & $6.3^{+0.9}_{-1.0}$ & $7.1^{+1.0}_{-1.1}$ & $3.5^{+0.5}_{-0.5}$ & $^{+2.1}_{-3.0}$ \\
$J-K>1.7$ & $493$ & $2.7 \pm 0.5$ & $9.0^{+0.9}_{-0.9}$ & $10.6^{+1.0}_{-1.0}$ & $4.8^{+0.4}_{-0.4}$ & $^{+2.6}_{-3.4}$ \\
$J-K>2.0$ & $381$ & $3.4 \pm 0.6$ & $10.2^{+1.0}_{-1.1}$ & $12.6^{+1.1}_{-1.2}$ & $5.3^{+0.5}_{-0.5}$ & $^{+2.9}_{-3.8}$ \\
$J-K>2.3$ & $267$ & $4.3 \pm 0.9$ & $11.1^{+1.3}_{-1.4}$ & $14.2^{+1.4}_{-1.5}$ & $5.8^{+0.6}_{-0.7}$ & $^{+2.8}_{-4.2}$ \\

& & & & & & \\

$R-K>2.9$ & $626$ & $1.5 \pm 0.4$ & $6.6^{+0.9}_{-1.0}$ & $7.6^{+1.0}_{-1.1}$ & $3.6^{+0.4}_{-0.5}$ & $^{+1.8}_{-2.4}$ \\
$R-K>3.4$ & $563$ & $2.3 \pm 0.4$ & $8.5^{+0.8}_{-0.9}$ & $9.9^{+0.9}_{-1.0}$ & $4.5^{+0.4}_{-0.4}$ & $^{+2.3}_{-3.0}$ \\
$R-K>3.9$ & $444$ & $3.2 \pm 0.5$ & $10.2^{+0.9}_{-1.0}$ & $12.4^{+1.1}_{-1.1}$ & $5.4^{+0.4}_{-0.5}$ & $^{+2.9}_{-3.7}$ \\
$R-K>4.4$ & $353$ & $4.7 \pm 0.7$ & $12.5^{+1.0}_{-1.1}$ & $15.5^{+1.2}_{-1.3}$ & $6.4^{+0.5}_{-0.5}$ & $^{+1.6}_{-4.3}$ \\

& & & & & & \\

$R<25$ & $341$ & $1.1 \pm 0.7$ & $5.4^{+1.6}_{-2.2}$ & $5.9^{+1.8}_{-2.3}$ & $3.0^{+0.8}_{-1.1}$ & $^{+2.3}_{-5.4}$ \\
$R>25$ & $303$ & $3.2 \pm 0.8$ & $10.0^{+1.4}_{-1.6}$ & $11.9^{+1.6}_{-1.7}$ & $5.2^{+0.7}_{-0.7}$ & $^{+3.0}_{-4.2}$ \\

& & & & & & \\

$K>19.3$ & $620$ & $1.6 \pm 0.4$ & $6.8^{+0.9}_{-1.0}$ & $7.7^{+1.0}_{-1.1}$ & $3.7^{+0.4}_{-0.5}$ & $^{+2.1}_{-3.0}$ \\
$K>19.7$ & $574$ & $1.7 \pm 0.4$ & $7.0^{+0.9}_{-1.0}$ & $7.9^{+1.0}_{-1.1}$ & $3.8^{+0.4}_{-0.5}$ & $^{+2.1}_{-3.0}$ \\
$K>20.1$ & $480$ & $2.1 \pm 0.5$ & $7.9^{+1.0}_{-1.1}$ & $9.0^{+1.1}_{-1.2}$ & $4.2^{+0.5}_{-0.5}$ & $^{+2.5}_{-3.4}$ \\
$K>20.5$ & $279$ & $3.3 \pm 0.9$ & $10.0^{+1.4}_{-1.6}$ & $12.3^{+1.6}_{-1.8}$ & $5.2^{+0.7}_{-0.8}$ & $^{+3.1}_{-4.4}$ \\

& & & & & & \\

$log(M)>10.4$ & $616$ & $1.4 \pm 0.4$ & $6.3^{+0.9}_{-1.1}$ & $7.1^{+1.0}_{-1.1}$ & $3.4^{+0.5}_{-0.5}$ & $^{+3.1}_{-4.4}$ \\
$log(M)>10.6$ & $543$ & $1.8 \pm 0.4$ & $7.2^{+0.9}_{-1.1}$ & $ 8.2^{+1.0}_{-1.1}$ & $3.9^{+0.5}_{-0.5}$ & $^{+2.2}_{-3.1}$ \\
$log(M)>10.8$ & $429$ & $1.7 \pm 0.6$ & $7.1^{+1.2}_{-1.4}$ & $8.2^{+1.4}_{-1.5}$ & $3.9^{+0.6}_{-0.7}$ & $^{+2.5}_{-3.5}$ \\
$log(M)>11.0$ & $325$ & $1.2 \pm 0.7$ & $5.9^{+1.8}_{-2.4}$ & $6.2^{+2.0}_{-2.5}$ & $3.3^{+0.9}_{-1.2}$ & $^{+2.4}_{-4.9}$ \\

\enddata
\tablenotetext{a}{All galaxies are selected using $K < 21$ unless otherwise specified}
\tablenotetext{b}{Estimated uncertainty in $r_0$ due to field to field variance.  See \S \ref{sec:simulations}}
\label{tbl:results}
\end{deluxetable}

\end{document}